\documentclass{aa}  
\usepackage{graphicx,natbib}
\usepackage{txfonts}
\usepackage{tablefootnote}
\usepackage{hyperref}
\newcommand{\teff}{\mbox{$T_{\mathrm{eff}}$}}
\newcommand{\teffp}{\mbox{$T_{\mathrm{eff,1}}$}}
\newcommand{\teffs}{\mbox{$T_{\mathrm{eff,2}}$}}

\newcommand{\loggp}{\ensuremath{\log g_{\mathrm 1}}}
\newcommand{\loggs}{\ensuremath{\log g_{\mathrm 2}}}
\newcommand{\vsini}{\mbox{$\upsilon \sin i$}}

\def\kms{$\mathrm{km\,s}^{-1}$}

\newcommand{\porb}{\mbox{\ensuremath{P_{\mathrm{orb}}}}}
\newcommand{\micron}{\mbox{$\mu m$}}

\newcommand{\rsun}{R\ensuremath{_\odot}}
\newcommand{\msun}{M\ensuremath{_\odot}}

\newcommand{\mstar}{\ensuremath{M_\star}}
\newcommand{\degree}{\mbox{\ensuremath{^\circ}}}   

\newcommand{\mprim}{\ensuremath{M_{\mathrm {1}}}}
\newcommand{\msec}{\ensuremath{M_{\mathrm {2}}}}
\newcommand{\rprim}{\ensuremath{R_{\mathrm {1}}}}
\newcommand{\rsec}{\ensuremath{R_{\mathrm {2}}}}
\newcommand{\rc}{R$_{\rm C}$}
\newcommand{\ic}{I$_{\rm C}$}
\newcommand{\vri}{VR$_{\rm C}$I$_{\rm C}$}

\begin{document}

   \title{Fundamental properties of the pre-main sequence eclipsing stars of MML~53 and the mass of the tertiary }

   \author{Y. G\'omez Maqueo Chew\inst{\ref{ia}}
          \and  L.\ Hebb\inst{\ref{hws}}
          \and  H.C.\ Stempels\inst{\ref{upp}}
          \and A. Paat\inst{\ref{noao}}
          \and K.G. Stassun \inst{\ref{vandy},\ref{fisk}}
          \and F. Faedi \inst{\ref{catania},\ref{warwick}} 
          \and R.A.\ Street \inst{\ref{lco}}
          \and G. Rohn \inst{\ref{cortland}}
          \and C. Hellier \inst{\ref{keele}}
          \and D.R.\ Anderson \inst{\ref{keele}}
          }

   \institute{
            Instituto de Astronom\'ia, Universidad Nacional Aut\'onoma de M\'exico, Ciudad Universitaria, 04510 Ciudad de  M\'exico, M\'exico  \label{ia}\\
                  \email{ygmc@astro.unam.mx}
             \and Department of Physics, Hobart and William Smith Colleges, Geneva, New York, 14456, USA\label{hws} 
             \and Department of Physics \& Astronomy, Uppsala University, Box 516, SE-75120 Uppsala, Sweden\label{upp}
             \and National Optical Astronomy Observatory, 950 N. Cherry Ave, Tucson, Arizona 85719, USA\label{noao}
             \and Department of Physics, Vanderbilt University, Nashville, TN, USA\label{vandy}
             \and Department of Physics, Fisk University, Nashville, TN, USA\label{fisk}
             \and INAF-Osservatorio Astrofisico di Catania, , Via S. Sofia 78, I-95123 Catania, Italy\label{catania}
             \and University of Warwick, Department of Physics, Gibbet Hill Road, Coventry, CV4 7AL, UK  \label{warwick}
             \and Las Cumbres Observatory Global Telescope Network, 6740 Cortona Dr. Suite 102, Goleta, CA 93117, USA  \label{lco}
             \and Physics Department, State University of New York at Cortland, Cortland, New York 13045, USA \label{cortland}
             \and Astrophysics Group, Keele University, Staffordshire, ST5 5BG, UK \label{keele}
             }
   \date{Received September 15, 1996; accepted March 16, 1997}

  \abstract
   {
   We present the most comprehensive analysis to date of 
  the Upper Centaurus Lupus eclipsing binary MML~53 (with P$_{\rm EB}=2.097892$~d), and for the first time, confirm the bound-nature of the third star (in a P$_{3}\sim9$~yr orbit) by constraining its mass dynamically. 
  Our analysis is based on new and archival spectra and time-series photometry, spanning 80\% of one orbit of the outer component. From the spectroscopic analysis, we determined the temperature of the primary star to be 4880 $\pm$ 100~K. The study of the close binary incorporated treatment of spots and dilution by the tertiary in the light curves, allowing for the robust measurement of the masses of the eclipsing components within  1\% (M$_1 = 1.0400 \pm 0.0067$ \msun\ and M$_2 = 0.8907 \pm 0.0058$ \msun),  their radii within 4.5\% (R$_1 = 1.283 \pm 0.043$ \rsun\ and R$_2 = 1.107 \pm 0.049$ \rsun), and the temperature of the secondary star (\teffs\ = 4379 $\pm$ 100~K). From the analysis of the eclipse timings, and the change in systemic velocity of the eclipsing binary and the radial velocities of the third star, we measured the mass of the outer companion to be 0.7~\msun\ (with a  20\% uncertainty). 
The age we derived from the evolution of the temperature ratio between the eclipsing components is fully consistent with previous, independent estimates of the age of Upper Centaurus Lupus (16$\pm$2~Myr).
At this age, the tightening of the MML~53 eclipsing binary has already occurred, thus supporting close-binary formation mechanisms that act early in the stars' evolution.
The eclipsing components of MML~53  roughly follow the same theoretical isochrone, but appear to be inflated in radius (by 20\% for the primary and 10\% for the secondary) with respect to recent evolutionary models. 
However, our radius measurement of the 1.04~\msun\ primary star of MML~53 is in full agreement with the independent measurement of the secondary of NP Per which has the same mass and a similar age. 
The eclipsing stars of MML~53 are found to be larger but not cooler than predicted by non-magnetic models, it is not clear what is the mechanism that is causing the radius inflation given that activity, spots and/or magnetic fields slowing their contraction, require the inflated stars to be cooler to remain in thermal equilibrium. 
     }

\titlerunning {MML~53: EB \& Tertiary}

 \maketitle


\section{Introduction}
  With the growing number of transiting planet surveys and follow-up radial velocity data, 
  the detection and characterization of eclipsing binary (EB) stars has seen a resurgence over the last decade. 
  Eclipsing binaries that are also double-lined spectroscopic systems have long provided crucial observational constraints 
  for stellar evolution models by allowing the direct measurements of the masses and radii of the components, and also importantly, a measure of 
  their temperatures \citep{Andersen1991,Torres2010,Stassun2014}.  Well-constrained stellar masses and radii are especially important for understanding pre-main-sequence (PMS) stars, 
  as these are rapidly evolving systems which have not yet fully contracted.
  
  Only ten years ago, the known pre-main sequence EBs with precisely measured properties were all members of the Orion nebula cluster, with ages between 1 and 2~Myr probing the youngest and earliest stages of stellar evolution, and of the Orion OB1 group, with an older stellar population of $\sim$10~Myr \citep[e.g.,][]{Mathieu2007}.  It has been over the last few years that new EB systems in other young clusters and associations have been discovered, observed and carefully analyzed. 
  Currently in the literature, there are only 14 known EB systems where the eclipsing components have directly measured masses (\mstar\ $\lesssim$ 1.4 \msun) and are on the pre-main sequence.
Of these, there are seven EB systems belonging to the Orion star formation complex:
  ASAS J052821+0338.5 \citep{Stempels2008}; 
  RX J0529.4+0041 \citep{Covino2001,Covino2004}; 
  V1174 Ori \citep{Stassun2004}; 
  Parenago 1802 \citep{Stassun2008,Cargile2008,GomezMaqueoChew2012}; 
  Parenago 2017 \citep{Morales2012}; 
  JW 380 \citep{Irwin2007}, and 
  2MASS J05352184--0546085 \citep{Stassun2006,Stassun2007,GomezMaqueoChew2009}. 
  There are five EBs that are members of the Scorpius-Centaurus OB complex:
  HD 144548 \citep{Kiraga2012,Alonso2015}; 
  MML 53 \citep{Hebb2010,Hebb2011}; 
  UScoCTIO 5  \citep{Kraus2015,David2016};
  EPIC 203710387 \citep{Lodieu2015,David2016}, and  
  EPIC 203868608 \citep{David2016}. 
  And there is one known EB in NGC 2264, 
  CoRoT 223992193 \citep{Gillen2014,Gillen2017}, 
  and one EB in the Perseus star-forming complex, 
  NP Per \citep{Perova1966,Lacy2016}. 
 Other pre-main sequence EB candidates have been identified but their fundamental properties have not yet been measured \citep[e.g.,][]{vanEyken2011,Morales2012}. 
 Given the large spread of measured masses (from brown dwarfs of $\sim$0.02~\msun\ up to $\sim$1.4~\msun\ stars) and radii ($\sim$0.25 to 2.4~\rsun) of the known eclipsing objects (see also Fig.~\ref{fig:mr}) and the spread in ages ($\sim$1--17~Myr) of their star-formation regions,  it is clear that analyses of these systems provides strong empirical constraints on models of pre-main sequence evolution of low-mass stars and brown dwarfs. 

MML~53, the first pre-main sequence EB discovered outside of the Orion star forming region and the subject of this paper, is an interesting pre-main sequence EB.
Its young, pre-main sequence nature  has been comfirmed by numerous observations measuring the X-ray emission, H$\alpha$ emission, and Li I $\lambda$6708 absorption from the system \citep{Wichmann1997a,Mamajek2002,Torres2006,White2007,Hebb2010}, and it has long been known as a spatial and kinematic member of the $16\pm2$~Myr old Upper Centaurus Lupus (UCL) subgroup of the Scorpius-Centaurus OB association \citep{Mamajek2002, Pecaut2016,Pecaut2012}.

The eclipsing nature of the system
was first discovered by \citet{Hebb2010} in data obtained as part of the WASP transiting planet survey \citep{Pollacco2006}.  Analysis of the 2006--2008 WASP light curve combined with additional radial velocity measurements taken in 2009 with the 1.5m telescope at Cerro Tololo Inter-American Observatory (CTIO) found the EB to be composed of
a $1.0$~M$_{\odot}$ and a 0.86~M$_{\odot}$ pair of stars in an 2.09~day eclipsing orbit \citep{Hebb2011}.
Features from a third unresolved star were also detected in the spectra.  As we show in this paper, variations in the  radial velocity of the tertiary component compared to the systemic radial velocity of the binary confirm MML~53 is a  gravitationally bound triple system.  In addition, \citet{Hebb2010} detected small changes of $\sim$3~minutes in the epoch of the eclipses from 2006--2008.   Subsequent WASP observations described in this work have confirmed these variations, which are attributed to light travel time effects.  
As the EB in MML~53 orbits the third star over the timescale of about a decade, the distance from the Earth to the EB changes causing the epoch of the eclipse minima to vary with its orbital position.

In summary, MML~53 is a 16~Myr old, hierarchical triple system consisting of a close eclipsing binary and a lower mass tertiary component that has recently been spatially resolved \citep{Schaefer2018}.  Due to its unique age among the known pre-main sequence EBs, precise measurements of the fundamental properties of its component stars have the potential to test a previously unconstrained part of parameter space in the theoretical stellar evolution models.  
In this paper, we present precise fundamental properties of the eclipsing components derived by incorporating new, high-quality spectroscopic and photometric observations of the MML~53 system into a comprehensive eclipsing binary model (\S\ref{sec:EBmodel}), which accounts for stellar surface spots and the effect of light from the tertiary star in the light curve modeling.  This paper also presents the first constraints on the mass and orbital parameters of the third stellar component through a combined
analysis of long term variations in the systemtic radial velocity of the EB and corresponding changes in its measured eclipses times.


\section{Observations}
\label{sec:observations}

In this section, we describe the new and archival observations utilized in our analysis of the MML~53 system for deriving the fundamental properties of its eclipsing stars and the orbital parameters of the bound tertiary component.  

\subsection{Photometric data}

\subsubsection{WASP photometry}

\begin{table}[h!]
\caption{MML~53 WASP light curve data}\label{table:lcswasp}
\begin{tabular}{l c c}
\hline\hline\\
HJD--2\,450\,000$^\dagger$ & $\Delta$mag & $\sigma_{\rm mag}$ \\
\hline\\
3860.38987~~~~~~~~~~~~~~~~~~~~~~~~ 	 & 	 0.0028 	 & 0.0723 \\ 
3860.39021 	 & 	 -0.0028 	 & 0.0625 \\ 
3860.39687 	 & 	 0.0076 	 & 0.0743 \\ 
3860.39731 	 & 	 0.0028 	 & 0.0743 \\ 
3860.40458 	 & 	 0.0011 	 & 0.0559 \\ 
  ... & & \\
\hline\\
\end{tabular}
\\
\footnotesize{$^\dagger$ Times are given in heliocentric Julian days (HJD) as produced by the WASP pipeline.}\\
\end{table}

\citet{Hebb2010} described the photometric time series data obtained on MML~53 between 2006--2008 as part of the WASP transiting planet survey \citep{Pollacco2006}.   Subsequently,
MML~53 was observed again in the field-of-view of the WASP-South telescope between 2011--2013.  All nights showing full or partial eclipses were extracted from the full WASP light curve and used to measure the epoch of minimum light of the eclipsing pair for each year of data between 2006--2013 as described in \S\ref{sec:epochtimes}.  
A total of 10328 photometric data points were obtained between February--August 2011; 10648 were obtained between February--June 2012 with an additional 23952 points in July and August 2012; and 76754 data points were observed using three cameras in an intensive campaign of this field between February--August 2013.  All WASP photometric data of MML~53 are provided in Table~\ref{table:lcswasp} (in full in online version\footnote{Tables 1--3 are only available in full in electronic form
at the CDS via anonymous ftp to cdsarc.u-strasbg.fr (130.79.128.5)
or via \url{http://cdsweb.u-strasbg.fr/cgi-bin/qcat?J/A+A/}}).  

\begin{figure*}[!h]
\centering
\includegraphics[width=1.0\hsize]{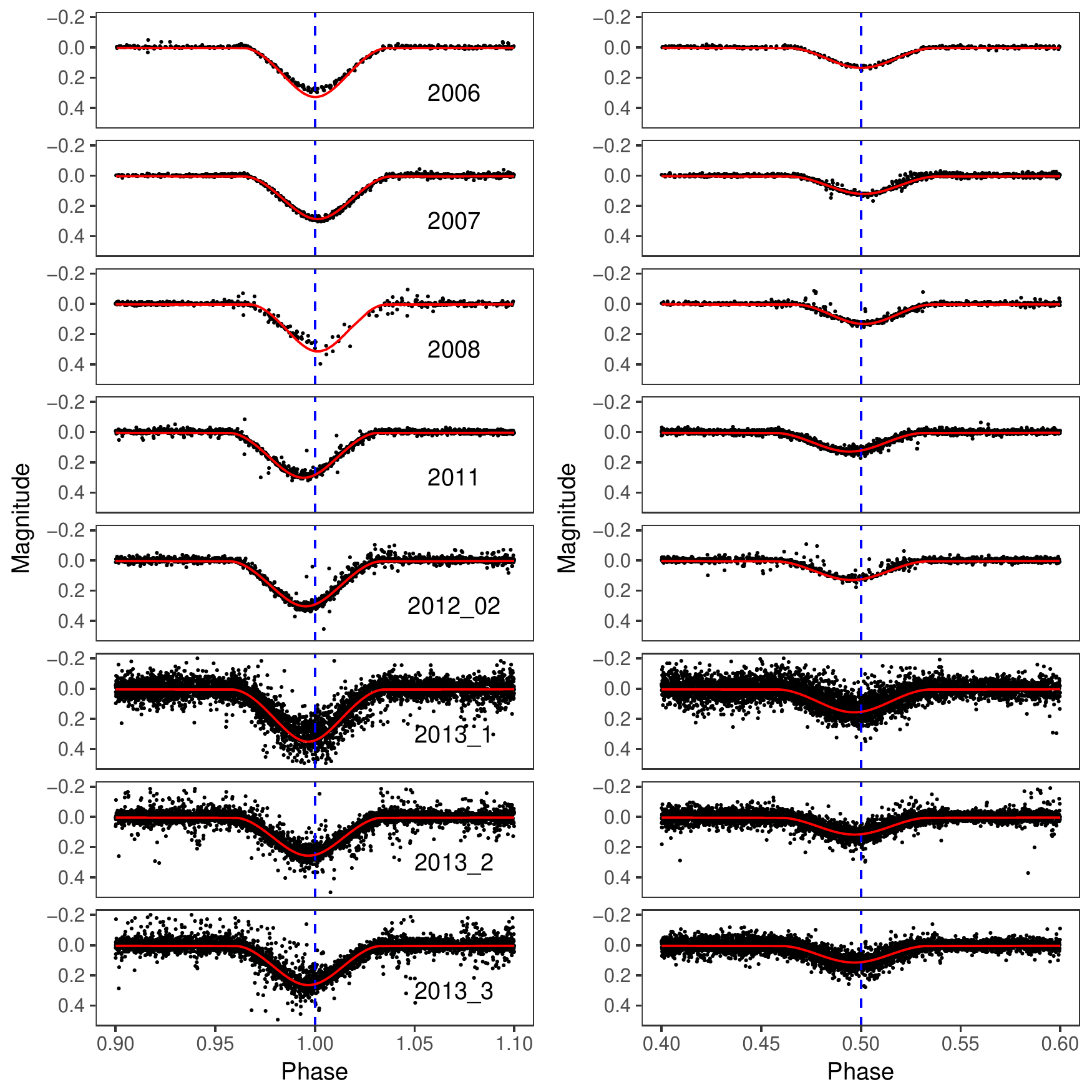}
\caption{Phase-folded WASP photometry of the primary and secondary eclipses of MML 53 from 2006-2013.  All data were phase-folded with the period of P=2.097892~$\pm$~0.000005~days, and time of minimum light in BJD$_{\rm TDB}$ units, T$_{0} = 2454972.650850$, derived from the detailed modeling of the 2009 CTIO data (marked at phase 1.0 with the vertical, dashed line).  Noticeable shifts in the time of eclipse minima due to the tertiary component are visible in these data.  Best fitting EB model light curves (described in \ref{sec:epochtimes}) are over plotted in red on all rectified light curves.   
These data are not used to derive the EB parameters, so the differing depths
between the model and observed eclipses do not affect the fundamental properties of the EB derived in this paper.
}
\label{fig:swaspphot}
\end{figure*}

These data were processed and removed of systematics with the standard WASP pipeline \citep{CollierCameron2006} resulting in 135078 brightness measurements obtained over this time period.   The typical photometric precision of the early 2006--2012 data is $\sim7$~mmag, which is measured by the standard deviation of the data points in the out-of-eclipse phases. 
After changing the WASP-South lenses from 200mm to 85mm in July 2012, the updated observing strategy lead to more observed data points with a lower precision of $\sim20$~mmag.  
Starspot modulations are present in these data from which we can measure a rotational period (\S\ref{sec:pre-eb}). However, the starspot modulation can affect the derived time of the eclipse epochs if they are not modeled correctly or removed. Therefore, each night of data was rectified by fitting a first or second order polynomial baseline to the out-of-eclipse data and subtracting the model values from the observed magnitudes at all times.  Individual out-of-eclipse data points were rejected at this stage if they deviate by more than 5$\sigma$ from the polynomial baseline.  The resulting phase-folded primary and secondary eclipses derived from the rectified light curves are shown in Fig.~\ref{fig:swaspphot} for each year of WASP data.   We omitted the July--August 2012 data since all eclipses occurred while the sun was up due to the near integer day period of the EB.   All data were phase-folded with the,  \porb~=~2.097892~$\pm$~0.000005~days, and time of minimum light in BJD$_{\rm TDB}$ units, T$_{0} = 2454972.650850$, derived from the detailed modeling of the 2009 CTIO data.  Noticeable shifts in the time of eclipse minima as compared to T$_{0}$ and due to the tertiary component are visible in these data.   The EB model light curves (described in \S\ref{sec:epochtimes}) are over plotted in red on all rectified light curves.  
The times of minimum light that change from year to year are defined with these EB models and are used to constrain the parameters of the tertiary's orbit in \S\ref{sec:M3}.

\subsubsection{Faulkes Telescope South photometry}
\label{sec:ftsphot}

\begin{figure}
  \centering
  \includegraphics[angle=0,width=\columnwidth]{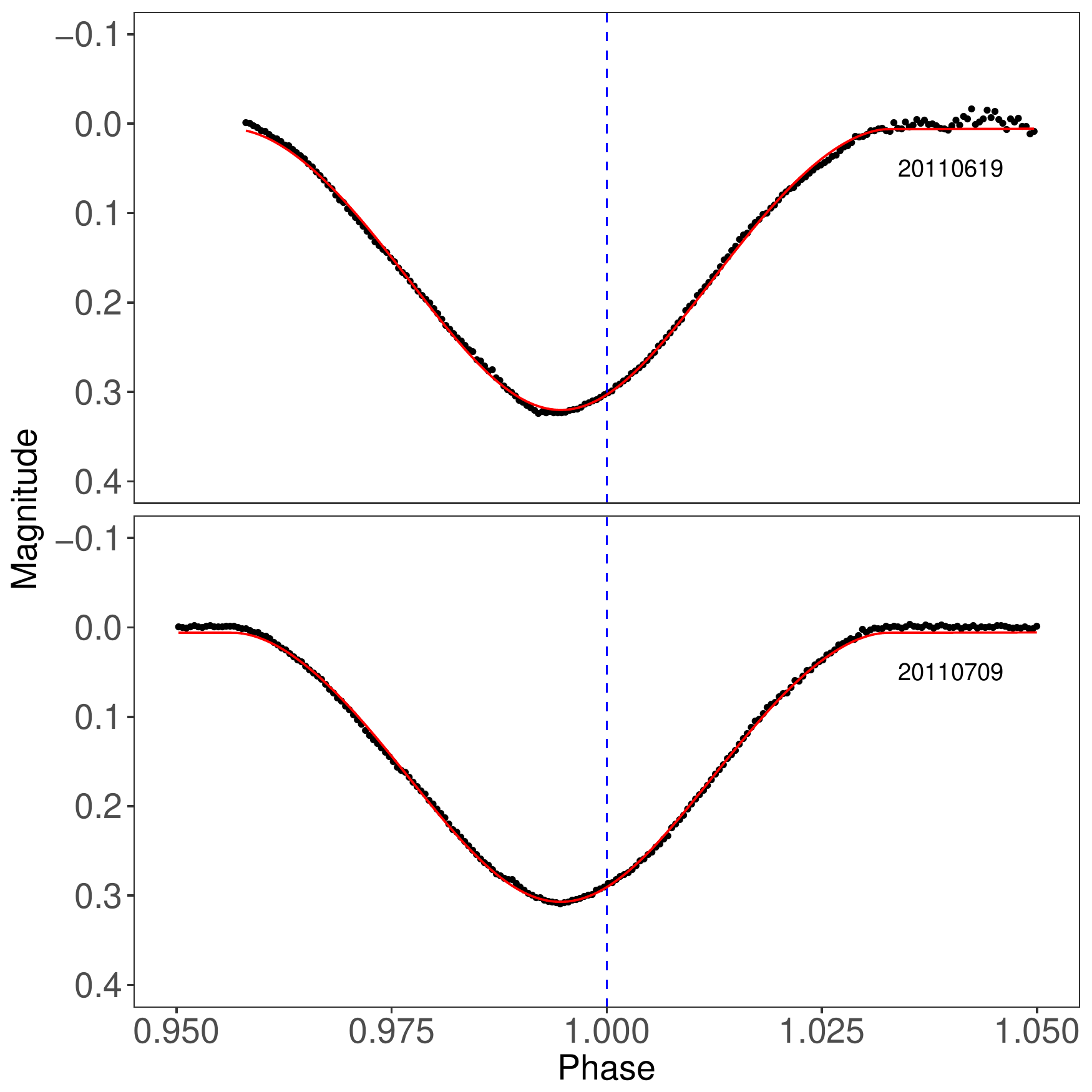}
  \caption{Differential $V$-band light curves of two primary eclipses of MML~53 taken on 19 June 2011 and 9 July 2011 with FTS. The data were converted to phase with the ephemeris derived from the detailed modeling of the 2009 CTIO data (\porb\ = 2.097892~$\pm$~0.000005~d) and time of minimum light in BJD$_{\rm TDB}$ units, T$_{0} = 2454972.650850$ shown with the vertical, dashed line. Noticeable shifts in the time of eclipse minima due to the tertiary component are visible in these data.  Best fitting EB model light curves (described in \ref{sec:epochtimes}) are over plotted in red on all rectified light curves.}
  \label{fig:ftsphot}
\end{figure}

Two primary eclipses of MML~53 were observed on 19 June 2011 and 9 July 2011 in order to continue tracking the eclipse timing variations.  The data were obtained with the Spectral Camera on the 2-m Faulkes Telescope South (FTS) through the Las Cumbres Observatory Global Telescope Network (LCOGT). 
We observed the target with 60~second exposure times repeatedly for approximately 5.5 hours in the Johnson V-band filter.  We employed the $2\times2$~binning mode for faster readout time and defocused the camera by 0.3~mm to avoid saturation.
The data were processed in the standard way with the LCOGT imaging data pipeline (BANZI)
\footnote{\url{https://lco.global/observatory/data/BANZAIpipeline/}}, which includes bad pixel masking, bias and dark frame subtraction, and flat-field division of each individual science frame with the best available calibration images.  The pipeline also performs
source extraction and astrometry.  The $5^{\prime}\times 5^{\prime}$ field-of-view of the instrument contained eight bright comparison stars that were used in deriving the differential magnitudes with a photometric precision of $\sim2$~mmag.  The phase-folded FTS light curves are shown in Fig.~\ref{fig:ftsphot}, and the data are given in Table~\ref{table:lcsfts}.

Attempts were made to get additional eclipse photometry in the 2012 and 2013 seasons, but poor weather prevented such observations.  

\begin{table}[h!]
\caption{MML~53 FTS V-band light curve data}\label{table:lcsfts}
\begin{tabular}{l c c}
\hline\hline\\
HJD$_{\rm UTC}$--2\,450\,000$^\dagger$ & $\Delta$mag & $\sigma_{\rm mag}$ \\
\hline\\
5731.99983~~~~~~~~~~~~~~~~ 	 & 	 -0.1372 	 & 	 0.0015 \\ 
5732.00075 	 & 	 -0.1346 	 & 	 0.0015 \\ 
5732.00169 	 & 	 -0.1331 	 & 	 0.0015 \\ 
5732.00265 	 & 	 -0.1295 	 & 	 0.0015 \\ 
5732.00360 	 & 	 -0.1287 	 & 	 0.0015 \\ 
  ... & & \\
\hline\\
\end{tabular}
\\
\footnotesize{$^\dagger$ Times are given in HJD$_{\rm UTC}$ as produced by the LCOGT pipeline.}\\
\end{table}

\subsubsection{CTIO photometry}

\begin{table}[h!]
\caption{MML~53 CTIO light curve data}\label{table:lcsctio}
\begin{tabular}{l c c c }
\hline\hline\\
BJD$_{\rm TDB}$--2\,450\,000 & $\Delta$mag & $\sigma_{\rm mag}$ & Filter \\
\hline\\
4970.47351 	 & 	 0.024 	 & 0.009 &  U \\ 
4970.48376 	 & 	 0.054 	 & 0.009 &  U \\ 
4970.49238 	 & 	 0.087 	 & 0.009 &  U \\ 
4970.49889 	 & 	 0.124 	 & 0.009 &  U \\ 
4970.50535 	 & 	 0.167 	 & 0.009 &  U \\ 
  ... & & \\
\hline\\
\end{tabular}
\end{table}

MML~53 was observed between May 18 and June 08, 2009 with the CTIO-1m telescope and
Y4K-Cam camera.  The detector consists of a 4K$\times$4K array of $15\mu$ pixels
placed at Cassegrain focus giving a $0.3^{\prime\prime}$/pixel platescale. Thus the entire
array projects to a $20^{\prime}\times20^{\prime}$ field of view.   The observed signal is fed into
four amplifiers causing the raw images to have a quandrant effect with the readnoise
between 11-12~$e^{-}$ and gain of 1.45-1.52~$e^{-}$/ADU, depending on the amplifier.
The detector has a readout time of 51 seconds and a
71k-electron well depth before non-linearity sets in. This converts to a
saturation of 40,000 counts/pixel in $1\times 1$ binning mode.

Throughout each observing night, MML~53 and the surrounding field
were monitored in the standard Kron-Cousins optical filter set (UBVR$_{\rm c}$I$_{\rm c}$) 
alternating continuously between all five filters.  
Exposure times were chosen to maximize the flux in the target star and nearby
reference stars while keeping the peak pixel value in MML~53 below $40,000$ counts.
The telescope was defocused to allow for longer exposure times to build up signal
in the fainter reference stars without saturating MML~53.
We adopted an exposure time of 7~seconds 
for the V, R$_{\rm C}$, and I$_{\rm C}$--band observations and longer exposures of 45~seconds and 90~seconds
in the B and U band filters, respectively, where the detector is less sensitive.   We achieved
an overall light curve cadence of approximately 8 minutes in each filter accounting for the exposure times, the read out time, and other overheads, like filter changes.
Since the orbital period of the eclipsing binary is very close to $\sim$2 days, three consecutive primary eclipses and three secondary eclipses were observed during the first six (6) clear nights of the observing campaign.  Four of these six nights were photometric.  On the other two nights, thin clouds were visible, but it did not affect the overall observing cadence or photometric precision of the data.    Due to poor weather, MML~53 was observed sparsely for the next seven nights (2009-05-24 to 2009-05-30).  The weather improved for the final week of the observing campaign, but all the eclipses occurred during the day, so these data only sample the out-of-eclipse variation and allow the characterization of the stellar spots (\S\ref{sec:spots}).  The eclipsing binary analysis described below uses only the first six (6) nights of data in which the eclipses occur. These data are presented in Table~\ref{table:lcsctio}. 

Flat field and bias calibration frames necessary for processing the images
were obtained during each observing night.  Sets of 11 bias frames were taken at the
beginning and end of each night, and single frames were observed periodically throughout each night. 
Eleven dome flats were observed per night in all five filters, and
twilight flats (3-4 per filter) were obtained on the few photometric nights in the beginning of the run.  Due to the relatively small number of twilight flats obtained in each filter, the dome flats were used
for the flat-field calibration correction.  

The images were processed in a standard way using routines written by L.\ Hebb in the IDL
programming language.  Each of the four amplifiers was processed independently.  All object and calibration
frames were first overscan corrected (by subtracting a line-by-line median overscan value), bias subtracted
and then trimmed.  Stacked bias images were created by averaging all bias frames observed each night and subtracted from all science  and flat-field frames.  All dome flats observed during the first six nights of the observing campaign were averaged into a single dome flat in each filter and then applied to the trimmed and bias-corrected science images.  

\begin{figure*}
\centering
\includegraphics[width=0.9\hsize]{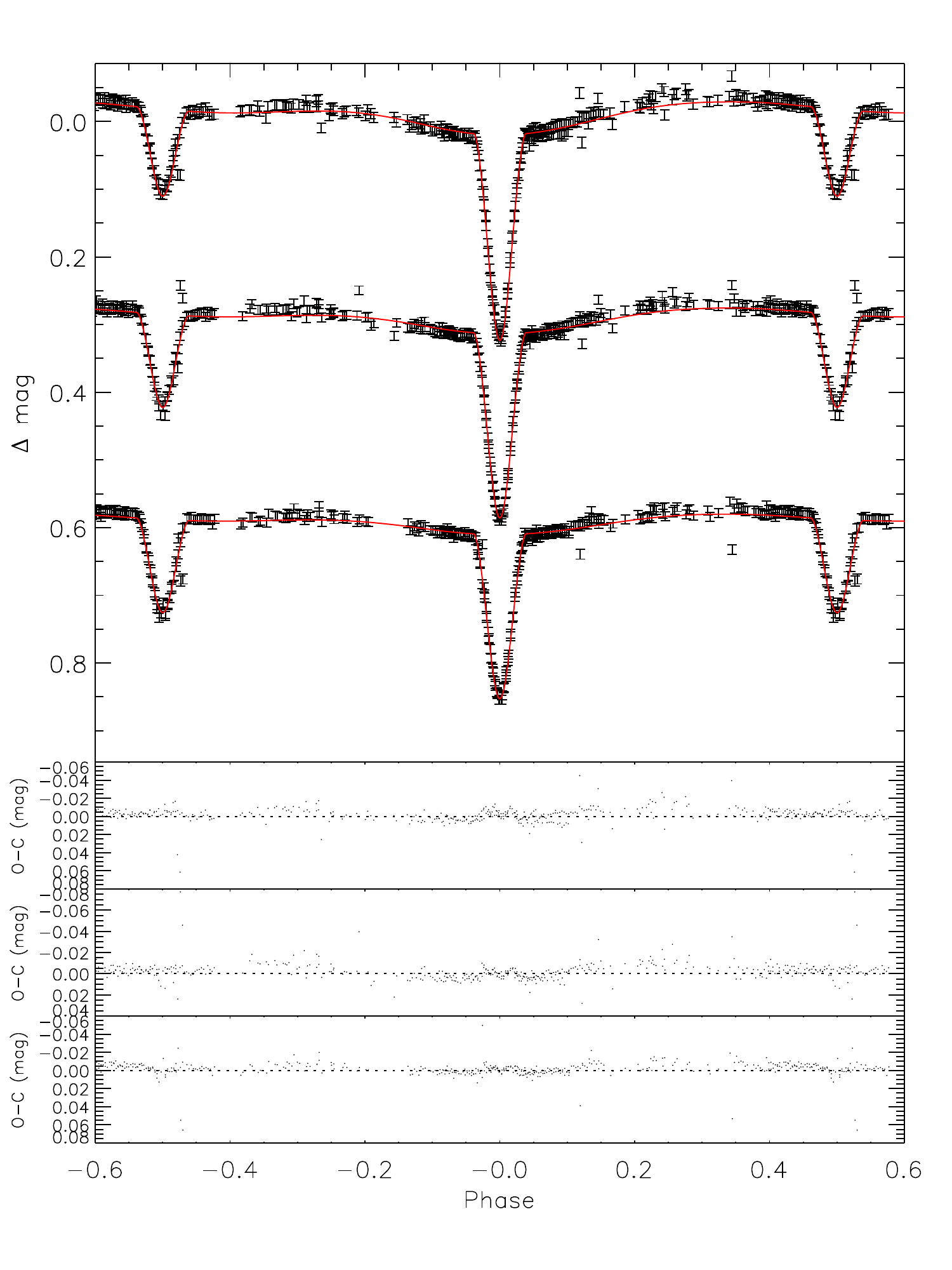}
\caption{CTIO VR$_{\rm C}$I$_{\rm C}$ photometry with the best-fit model light curves. 
On the top panel, we show the CTIO time-series photometry of MML 53, from the top are the V--, R$_{\rm C}$--, and I$_{\rm C}$--band differential photometry measurements shown by the black points with individual error. 
The light curves were arbitrarily separated in $\Delta$mag for clarity. The model light curves corresponding to the final solution, including third light and stellar spots (\S\ref{sec:spots}), are shown by the
continuous red lines. The three bottom panels show the residuals to the best fit model for each of the light curves, V, R$_{\rm C}$, and I$_{\rm C}$, respectively from the top. Our solution is able to reproduce well the duration and depth of the eclipses in the different bands and the variation attributed to spots. 
The r.m.s.\ in the residuals in each filter ($\sim$8 mmag) are comparable to the errors in the photometric measurements. 
}
\label{fig:lcs}
\end{figure*}

\begin{figure}[!h]
\centering
\includegraphics[width=1.0\hsize,height=0.25\textheight]{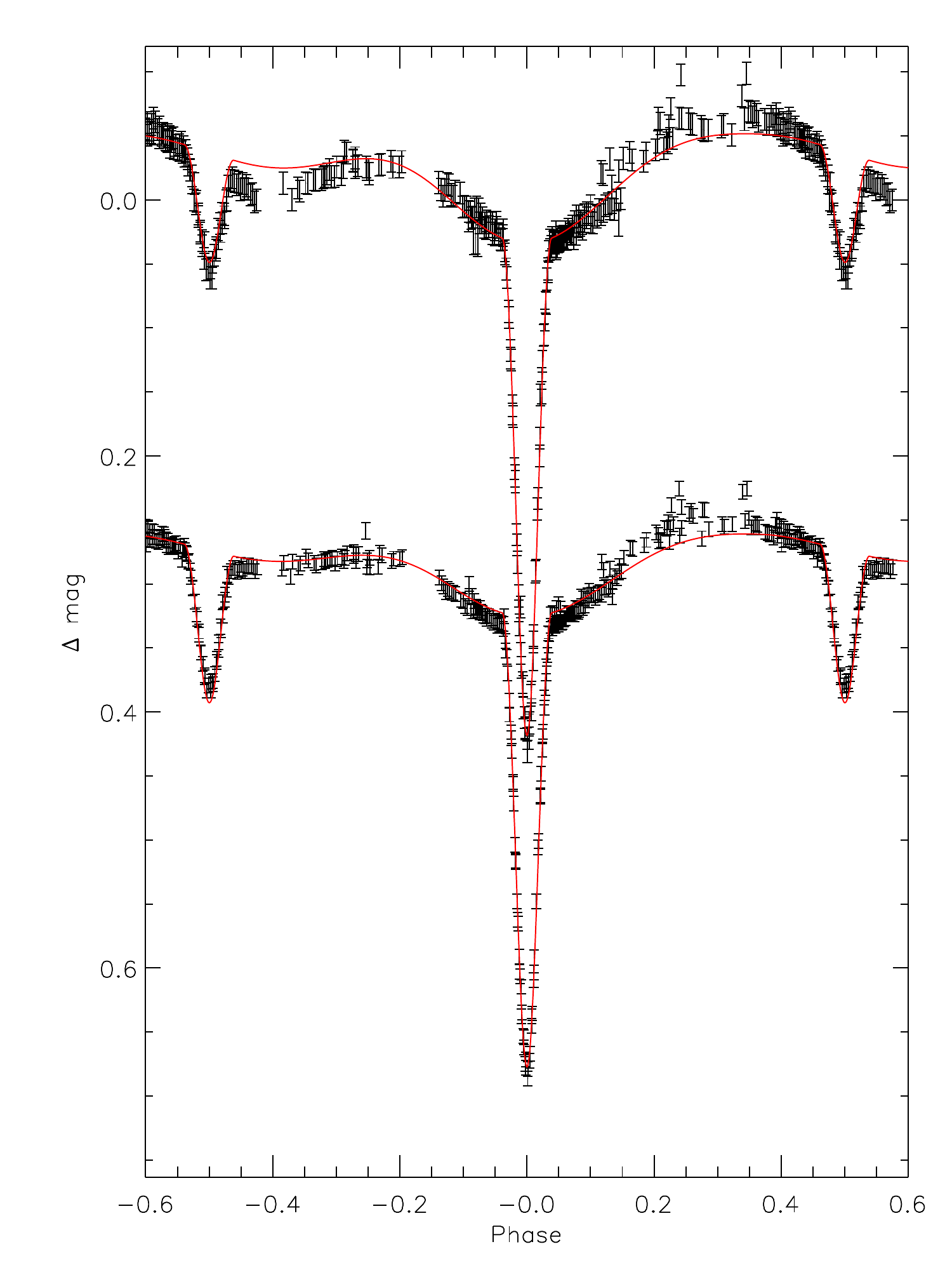}
\caption{CTIO UB photometry with the best-fit model light curves. 
We show for reference the acquired photometry in the UB-filters in black data points. The model light curves shown by the red continuous lines are the best-fit model to the \vri\ light curves and RV curves, fitting the third light in the U and B bands to fit the data. 
}
\label{fig:ublcs}
\end{figure}

Souce detection and aperture photometry were performed on all processed science images using the 
Cambridge Astronomical Survey Unit catalog extraction software \citep{IrwinLewis2001}. 
The  software  has  been  compared  with  SExtractor \citep{Bertin1996}
and found to be very similar in the completeness, astrometry and photometry tests
\footnote{\url{https://www.ast.cam.ac.uk/ioa/research/vdfs/docs/reports/simul/index.html}}.
This photometry software was applied to all processed images of MML~53.  Adopting conservative parameters
to define the detection threshold, the target star and dozens of fainter stars in the field were detected in each image.  Aperture photometry was performed on all detected stars using a 4~pixel radius circular aperture, which was selected to match the typical seeing over the first six nights of the observing run. 
The same aperture was used on all nights of data.  Eight bright, non-variable reference stars were selected from the many detected stars and used to perform differential photometry on the target star.  In each image, the flux from all reference stars was summed into a single {\it super} comparison star that was divided by the aperature flux from MML~53 and converted to a differential magnitude.  The resulting phase folded differential photometry light curves of MML~53 obtained from the first six nights of the observing run are shown in Fig.~\ref{fig:lcs} (VR$_{\rm C}$I$_{\rm C}$) and Fig.~\ref{fig:ublcs} (UB).

\subsection{Spectroscopic data}

The spectroscopic data presented in this paper are used to model the short term radial velocity variations of the primary and secondary EB components and to track the long term secular variations in velocity as the EB and the tertiary orbit their common center of mass.     

\subsubsection{UVES spectra}

MML~53 was observed fourteen times between 14 July 2009 and 18 September 2009 with the UVES spectrograph on the ESO Very Large Telescope (program ID 383.C-080).
The observations were obtained with the dichroic mode on the instrument, with the blue arm centered at 3900{\AA}, and the red arm at 5800{\AA}. Data in the blue arm are of poor quality and were not  considered in this paper. We adopted a slit width of 0.6$^{\prime\prime}$ which allows for achieving a resolution, R$\sim 60,000$ at the red end of the spectrum. With exposure times of 3~minutes per observation, we achieved a signal-to-noise of 100 on the V$\sim 10.8$ star. The data were processed with the REDUCE package \citep{Piskunov2002}, which uses advanced order-tracing and slit-modeling techniques to reconstruct and extract the stellar spectrum.

\subsubsection{FEROS spectrum}  
One high resolution ($R\sim50,000$) spectrum of MML~53 covering a wavelength range between 3765 and 8862 \AA\ was found in the European Southern Observatory (ESO) archive.  The spectrum was obtained on 23 June 2006 using the FEROS \'echelle spectrograph on the 2.2m MPG/ESO telescope. This spectrum was presented in \citet{Hebb2010} where it was used to confirm the presence of the tertiary star and measure the radial velocity values of all three components.   The details of the radial velocity analysis can be found in that paper, but the measured velocities are $-85.8, 111.1$, and $-3.5$~km~s$^{-1}$, for the primary secondary and tertiary, respectively.  Based on our experience, we adopted an uncertainty of 1.1~km~s$^{-1}$ for the an individual radial velocity (RV) measurement from this instrument.  Using the mass ratio derived from the final EB analysis (\S\ref{table:eb}) and these primary and secondary star RV measurements, we determineds the systemic radial velocity for the EB to be $5.2 \pm 0.8$~km~s$^{-1}$ at the time of this observation.  We report this value in the Table~\ref{tab:systemicRVs} and use it in the binary-tertiary analysis (\S\ref{sec:M3}).

\subsubsection{CTIO spectra}
A series of thirteen spectra of MML~53 were obtained in queue mode between 18 May 2009 and 12 June 2009 with the SMARTS 1.5m \'echelle spectrograph\footnote{See
\url{http://www.ctio.noao.edu/~atokovin/echelle/index.htm}.} at the
Cerro Tololo Inter-American Observatory (CTIO).  We also observed a single spectrum with the same instrument the following season on 09 September 2010 to continue monitoring the radial velocity variations in the tertiary star.   
A detailed description of the processing and analysis of the 2009 data are presented in \citet{Hebb2011}, which we summarize briefly here since it is the same for the newly presented 2010 spectrum.  

The bench-mounted spectrograph has a fixed cross-disperser and \'echelle grating, but accommodates a variety of slit widths that allow for resolutions of 25,000--40,000.  In order to maximize the signal-to-noise in these observations, we obtained $3\times 600$s exposures each night with a large slit width of 140 \micron\, which translates into a signal-to-noise of $S/N\sim$25 per resolution element and a resolution of R$\sim$25,000, which is sufficient to identify and resolve the three individual components of MML~53.  
The spectral images taken on each night were processed in the standard way with overscan subtraction, 2-D bias subtraction, trimming, and flat-fielding before the three individual images were median combined.  The spectra were extracted from each processed science frame and then wavelength calibrated with nightly ThAr lamp exposures using standard \'echelle data processing routines in IRAF\footnote{IRAF is distributed by the National Optical Astronomy Observatories, which are operated by the Association of Universities
for Research in Astronomy, Inc., under cooperative agreement with the
National Science Foundation \citep{Tody93}.}.  A single radial velocity standard was obtained on each night of the science observations and processed in an identical fashion.  In 2009, a single spectrum of HD~81797 was used as the radial velocity template in the cross-correlation analysis, and in the 2010 season, a single 60~second exposure of HD~223807 was observed for the same reason. This star has a radial velocity of $-15.83$~\kms \citep{Nidever2002}.  

A cross-correlation analysis using the IRAF routine {\sc fxcor} was performed on the calibrated spectra obtained in 2009 to measure twelve and ten independent radial velocities for the primary and secondary components, respectively.  This analysis, presented in \citet{Hebb2011}, resulted in measurements of the mass ratio and the systemic radial velocity of the EB of $+1.4 \pm 0.9$~km~s$^{-1}$.   During this time, the radial velocity of the tertiary star was also measured using {\sc fxcor} in five spectra obtained near quadrature.  The average radial velocity of the tertiary derived from these spectra is $+11.0 \pm 3.0$~km~s$^{-1}$.
In this paper, we analyzed the reduced spectrum from 2010 as described in \S\ref{sec:LSD} and derived the radial velocity of all three components.

\begin{figure}
\centering
\includegraphics[width=1.0\hsize]{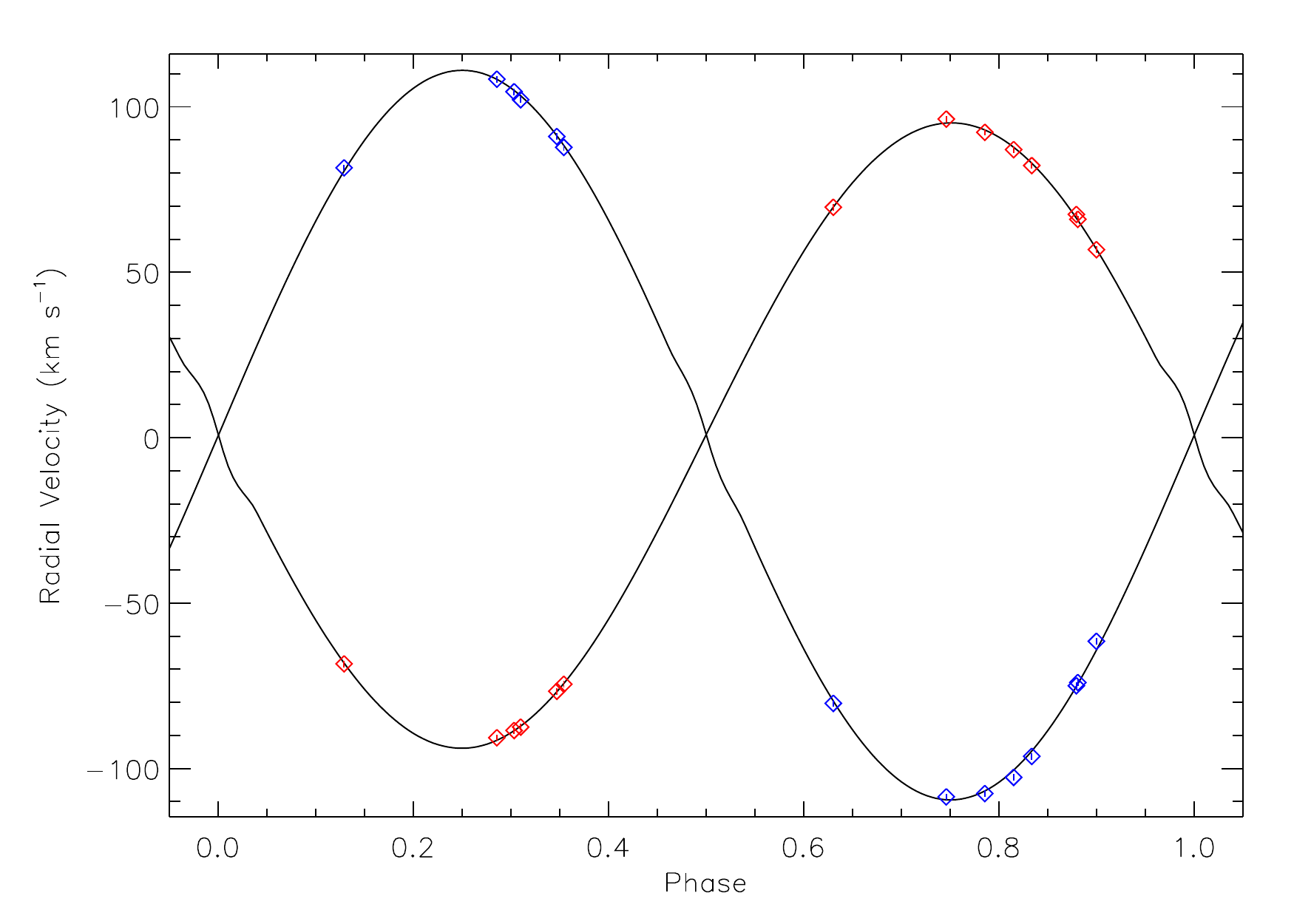}
\includegraphics[width=1.0\hsize,height=0.15\textheight]{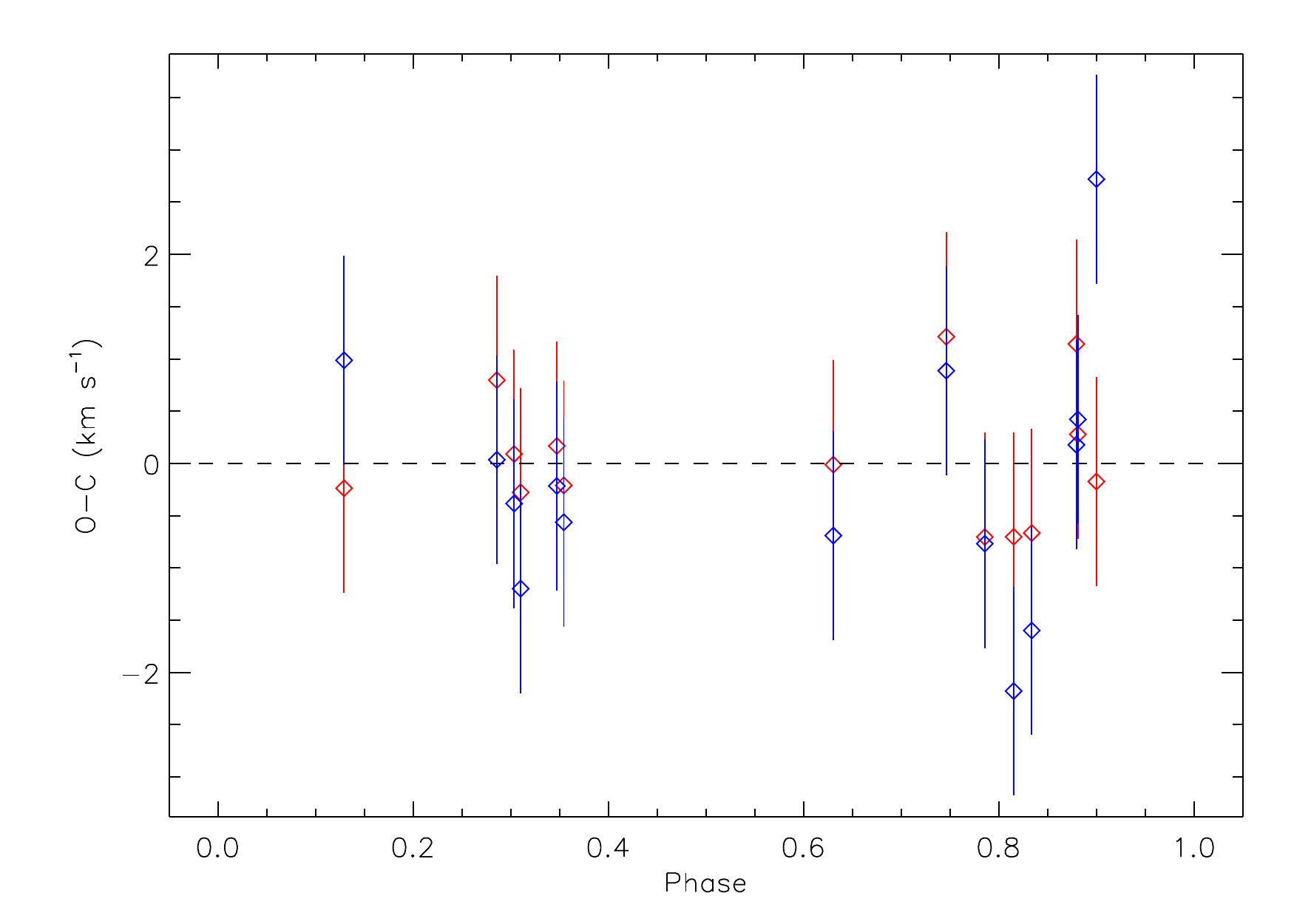}
\caption{UVES radial velocity curves and best-fit model. 
{\bf Top}: We show the radial velocity measurements from the UVES data. The red points correspond to the primary RV measurements, and the blue points to the secondary RVs. The errors in the RV measurements are also shown, but are smaller than the data points. The continuous black lines correspond to the best-fit model of the Keplerian orbit of the eclipsing components.  {\bf Bottom:} We show the residuals to the fit in red for the primary and in blue for the secondary, and the uncertainties that correspond to the error in each RV measurement.  The r.m.s.\ in the residuals are comparable to the errors in the RV measurements, $\sim$0.6 \kms\ for the primary and $\sim$1.1 \kms for the secondary.}
\label{fig:rvs}
\end{figure}

\section{Analysis and results}

The various analysis steps to characterize this system are not independent.  First, the determination of the mass of the tertiary body described in Sect.~\ref{sec:M3} depends on knowing the sum of the masses in the eclipsing binary ($M_{\rm B}$ = \mprim\ + \msec) 
which is derived from the EB model described in Sect.~\ref{sec:EBmodel}.  However, the final EB model solution depends on knowing the value of the third light, and the third light depends on the mass of the tertiary.  Furthermore, the spectral disentangling (described in Sect.~\ref{sec:LSD}) requires the relative luminosity ratios of the three unresolved components of the system which depends on the EB model and the binary-tertiary model.  Finally, the spectral synthesis necessary to determine the temperature of the primary star requires knowledge of the gravity of the primary and secondary stars determined from the EB model.  Therefore, the analysis steps described below were performed in an iterative manner until all solutions were consistent with each other, and the derived properties are measured robustly.  Below, we describe the details of each analysis step and the final results derived from it during the final iteration.

\subsection{Spectroscopic analysis}
\label{sec:LSD}

The high-resolution UVES spectra obtained in 2009 spectrally resolve all three components of MML~53, and cover a range of orbital phases of the system. This allowed us to determine the individual radial velocities of the components, as well as to perform a spectroscopic analysis of the three stars in this system.

    \begin{figure*}
    \centering
    \includegraphics[width=1\linewidth,trim=3cm 2.1cm 2.2cm 2.2cm, clip=true]{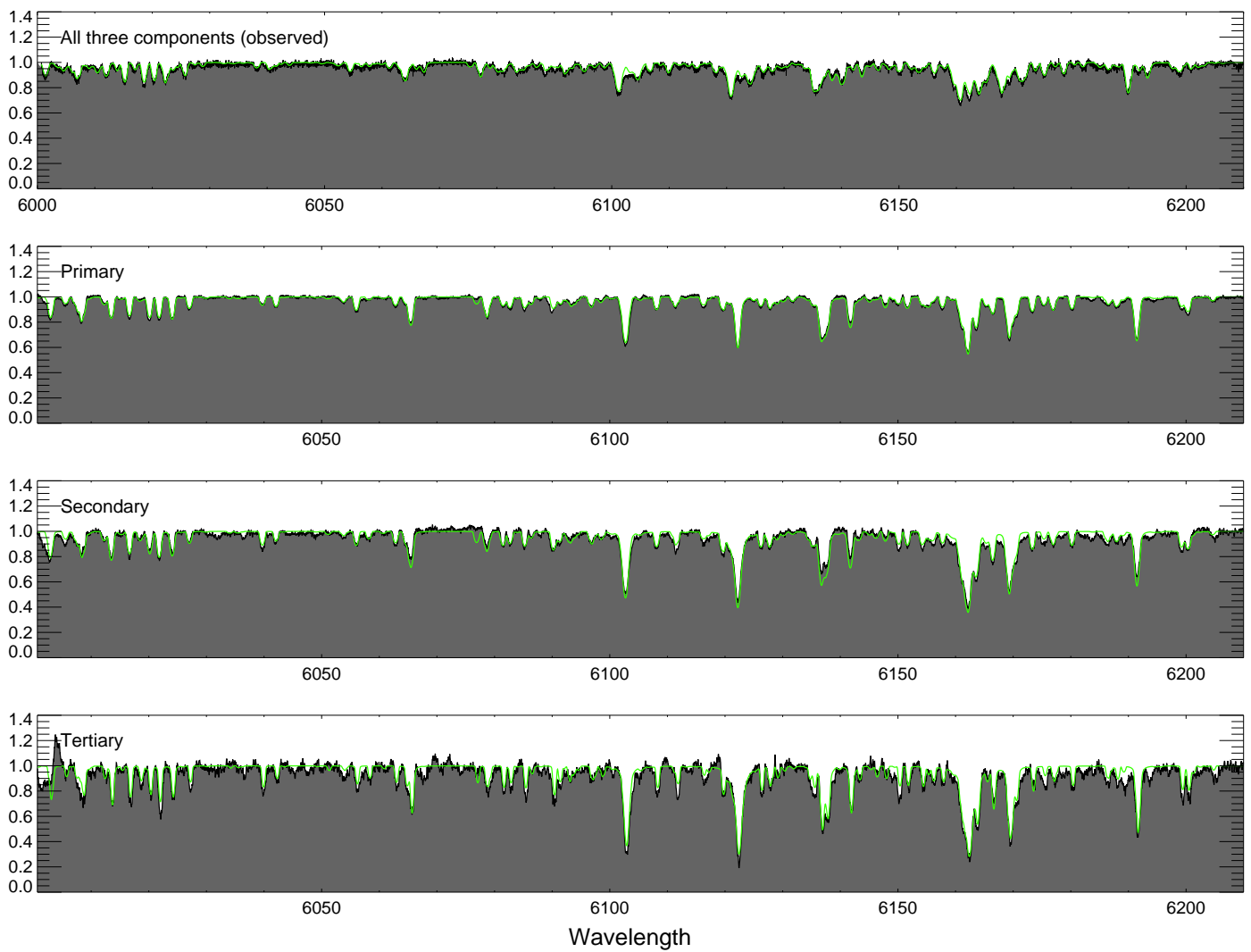}
    \caption{
    Observed spectrum of MML~53 and model spectra of all three stellar components. 
    This figure illustrates the agreement between the observed spectrum of MML~53 (black solid line and gray underlying area) and synthetic model spectra (green) for each of the three stars in the system. The top panel shows the spectrum as observed on 18 august 2009, when the absorption lines of this triple-lined spectroscopic system were well separated. The bottom three panels show the disentangled spectra of each individual stellar component. The synthetic spectrum in the top panel is a luminosity-weighted combination of the synthetic spectra shown in the lower panels, taking into account the radial velocity offset of each component.  
 }
      \label{fig:disentangling}%
     \end{figure*}

Using the method of least-squares deconvolution 
\citep[LSD, see][]{Donati1997,Kochukhov2010} we determined from each spectrum combined-average line profiles of MML 53, concentrating on the region between 5500{\AA} and 6500{\AA} which is populated by a large number of narrow absorption lines. We recovered three profiles one for each star, which were separated in velocity-space. 
To each of the recovered profiles we then fitted a three-component model consisting of three rotational profiles, calculated by disk-integrated radiative transfer, which includes limb-darkening and non-rotational broadening processes such as micro- and macroturbulence. In this analysis, we used a macroturbulence value of 1.2~\kms, and a microturbulence value of 1.0~\kms, as appropriate for pre-main sequence stars \citep[e.g.,][]{Padgett1996}. This allowed us to determine radial and projected rotational velocities for each component. The recovered radial velocities 
were measured relative to synthetic spectra, which are based on laboratory wavelengths from VALD \citep{Piskunov1995,Kupka1999}, and 
are presented in Table~\ref{tab:rv_lsd}, and used in Sects.~\ref{sec:pre-eb} and \ref{sec:EBmodel}. The measured rotational velocities (\vsini) are $30.6 \pm 1.0$ \kms, $26.6 \pm  1.3 $ \kms, and $25.8 \pm 3.2$ \kms\ for the primary, secondary and tertiary components, respectively. The errors in the measured \vsini\ were derived from the standard deviation of the \vsini\ derived from each of the UVES spectra (0.3, 0.8, and 3~\kms, for the primary, secondary and tertiary respectively) to which we  added in quadrature 1.0~\kms uncertainty  to account for systematic errors due to the continuum normalization and/or macroturbulence and microturbulence values used in the analysis.

We also analyzed the reduced CTIO spectrum from 2010 using LSD and find the radial velocity of all three components to be $-92$, $103$, and $18$~km~s$^{-1}$, for the primary secondary and tertiary, respectively.  An uncertainty of $2.5$~km~s$^{-1}$ was adopted for these measurements. Using the mass ratio derived from the final EB analysis (\S\ref{table:eb}) and these primary and secondary star RV measurements, we determine the systemic radial velocity for the EB to be $-2.0 \pm 1.8$~km~s$^{-1}$ at the time of this observation.  We report this value in the Table~\ref{tab:systemicRVs} and use it in the binary-tertiary analysis (\S\ref{sec:M3}).   

We also applied the method of spectral disentangling \citep[see][]{Bagnuolo1991} to our set of UVES spectra. This technique inverts the relation that each observed spectrum is a linear combination of the spectrum of each of the three stellar components. Using the radial velocities recovered above as input we numerically reconstructed the spectra of the three components as described in \citet{Stempels2011}. Since this technique requires an assumption of the relative luminosity ratio of the three components, we adopt the values presented in Table~\ref{table:thirdlight} for this parameter. The recovered spectra are typical of young K-type dwarfs, with Li {\sc i} 6708\AA\ absorption clearly present. Both the secondary and the tertiary components are affected by narrow emission in H$\alpha$ and He {\sc i} 5876{\AA}; no emission was present in the spectrum of the primary star, although the H$\alpha$ line appeared to be filled-in. Because of the composite spectra, a more detailed analysis of the activity of the MML~53 stars is beyond the scope of this paper.

Once extracted, we calculated synthetic spectra for each component with the SME software package \citep{Valenti1996,Piskunov2017}, using MARCS model atmospheres \citep{Gustafsson2008} and atomic and molecular line lists from the VALD database \citep{Piskunov1995,Kupka1999}. While it is possible to estimate the surface gravity ($\log g$) from the spectrum, this parameter is much better determined from the masses and radii recovered from the EB modeling (see \S\ref{sec:EBmodel}), and these values were therefore used as an input value when determining the effective surface temperatures \teff\ for the eclipsing stars.   
From this analysis we recover for the primary, secondary, and tertiary \teff\ = 4880~$\pm$~100 K, \teff\ = 4482~$\pm$~100 K, and \teff\ = 4500~$\pm$~250 K, respectively.  
In the calculation of synthetic spectra, the metallicity was assumed to be solar, and the micro- and macroturbulence were estimated to be $1.0$ and $1.2$~\kms. 

The agreement of observed and disentangled spectra with synthetic spectra based on these parameters is illustrated in Fig.~\ref{fig:disentangling}.  In our analysis we find that the overall agreement between the synthetic and observed spectra is excellent for the primary star.  Thus, we are highly confident in the primary star parameters derived from this synthesis.  However, for the secondary star, we required to include an additional continuum source caused by magnetic activity, also referred to as ``veiling''.  Without the veiling corresponding to 20\% of the light of the secondary star, 
the depth of calculated Na~D lines are consistently too deep, while the shape (which is highly temperature-dependent) corresponds well to the observed spectrum. Also,  H$\alpha$ emission suggests significant levels of magnetic activity is present in both the secondary and tertiary spectra.  This unknown veiling quantity can have a moderate affect on the derived parameters.  The tertiary disentangled spectrum is also in good agreement with the observation, however the properties are more uncertain because the luminosity of the tertiary is much lower than the other two components (as reflected in the uncertainty in the derived effective temperature). 

Fortunately, only the primary star temperature and \vsini\ are necessary (along with the EB model solution) to derive all individual properties of the binary components, which is the aim of this paper.

\begin{table}
\caption{Radial velocity measurements}\label{tab:rv_lsd}
\begin{tabular}{l c c c c c}
\hline\hline\\
Instrument	        & BJD$_{\rm TDB}$      & Primary       & Secondary      & Tertiary   \\
\phantom{Instrument} & $- 2\,450\,000$ & RV  & RV   &  RV    \\
\phantom{Instrument} & \phantom{JD} & (kms$^{-1}$) & (kms$^{-1}$) & (kms$^{-1}$) \\
\hline\\
FEROS$^a$    & 3909.62035  &    -85.8  &  111.1  & -3.5 \\   
\hline\\
UVES & 5026.66177  &      96.3$^b$    &    -108.6$^c$   &    11.1$^d$ \\
     & 5040.51000  &     -76.7    &      91.0   &     7.8 \\
     & 5042.47920  &     -90.7    &     108.3   &    10.4 \\
     & 5044.61337  &     -88.5    &     104.6   &     9.1 \\
     & 5045.62608  &      92.3    &    -107.6   &     10.1 \\
    &  5049.49566  &      69.6    &     -80.3   &     10.3 \\
     & 5060.55105  &      56.8    &     -61.6   &    9.0 \\
     & 5061.50408  &     -74.5    &      87.7   &     8.2 \\
     & 5062.47025  &      87.1    &    -102.7  &    11.1  \\
     & 5081.48640  &      67.5    &     -75.0   &    11.2 \\
     & 5081.49009 &      66.0    &     -74.0   &    11.6 \\
     & 5083.48843 &      82.2    &     -96.3   &    11.5 \\
     & 5084.48765  &     -87.5    &     102.1   &    10.3 \\
     & 5092.49966  &     -68.4    &      81.6   &     8.9 \\
\hline\\ 
CTIO 2010$^e$ &  5449.48870  &    -92  &  103  & 18 \\   
\hline
\end{tabular}
\footnotesize{$^a$ Uncertainty of 1.1 km s$^{-1}$ for all three components. \\
$^b$ Uncertainty of 0.6 km s$^{-1}$ derived from the residual scatter relative to the model.\\
$^c$ Uncertainty of 1.1 km s$^{-1}$ derived from the residual scatter relative to the model.\\
$^d$ Uncertainty of 0.5 km s$^{-1}$ from the scatter in the measurements.\\
$^e$ Uncertainty of 2.5 km s$^{-1}$ for all three components.
}
\end{table}

\begin{table}
\caption{Systemic EB and tertiary radial velocities}
\label{tab:systemicRVs}    
\centering                       
\begin{tabular}{l c c c }
\hline\hline\\
Instrument     & BJD$_{\rm TDB}$  & $\gamma_{EB}$ & RV$_3$   \\
   +Epoch  &   $- 2\,450\,000$ &     (\kms)    &      (\kms)    \\
\hline \\
FEROS 2006 & 3909.62035   &  $5.2 \pm 0.8$  & $-3.5  \pm  1.1 $ \\
UVES 2009  &  5061.24102  &  $0.76 \pm 0.15$  &  $10.1 \pm  1.3 $\\
CTIO 2009  &  4982.22138  &  $1.4 \pm 0.9$   & $11.0  \pm  3.0 $\\
CTIO 2010  &  5449.48870  &  $-2.0 \pm 1.8$   &  $18.0  \pm  2.5 $ \\ 
\hline
\end{tabular}
\end{table}

\subsection{Preliminary eclipsing binary model}\label{sec:pre-eb}

For the analysis of the eclipsing components, we utilized the information from previous studies of the MML 53 system and adopt the assumptions described in this section. 

The orbital period of the eclipsing binary was adopted from the careful determination presented in the discovery paper from the detailed analysis of the times of the eclipses \citep{Hebb2010}.

Utilizing all the available radial velocity data and light curves, we explored the possibility of a non-circular orbit. None of the solutions that allowed for a non-zero eccentricity converged. Moreover, the sinusoidal shape of the RV curves and the fact that the primary eclipse occurs at phase 0.0 and the secondary at phase 0.5  are robust indicators that the orbit is circular. Thus, for the rest of this analysis we adopt a circular orbit for the eclipsing binary (with eccentricity $e$ = 0.0).

We applied a Lomb-Scargle periodogram to each independent WASP light curve after removing the primary and secondary eclipses. Searching for periods between 0.5--30~d resulted in seven of the nine light curves having the strongest peak close to the same rotation period. We averaged the seven independent periods to determine a rotation period of 2.091 $\pm$ 0.013~d from the spot modulation present in the WASP light curves. 
Fully consistent with the orbital period, the components are found to be rotationally synchronized to their orbital motion, as expected for a circular orbit given that the tidal circularization timescale is expected to be longer than the synchronization timescale \citep[e.g,][]{Zahn1977,Mazeh2008}. 

Given that the spin-orbit alignment timescale is of the order of the
synchronization timescale, we also assume that the stellar spin axes are
aligned with the plane of the eclipsing binary orbit. Calculating the condition producing spin-orbit misalignment in the inner binary due to a tertiary component from \citet{Anderson2017}, we find that in the case of MML 53 the eclipsing components of the binary are not likely to be misaligned.

Given that the UVES RV curves were taken over a period of $\sim$66 days (corresponding to $\sim$2\% of the tertiary orbit) and that the peak-to-peak RV variation of the center of mass velocity of the EB ($\gamma_{\rm EB}$) is $\sim$10 \kms, we considered the change in $\gamma_{\rm EB}$ to be negligible over the timespan that the UVES observations encompass. 

Because the \citet{Baraffe2015} stellar evolutionary models do not provide constraints in the U and B broadband filters, and thus, there are no constraints on the third light in those bands, we did not use the UB light curves to derive physical properties of the eclipsing components. 

Only in this preliminary model, we adopted the effective temperature of the primary component from the previous spectroscopic determination \citep[\teffp = 4890~K;][]{Hebb2010}. The primary temperature was updated from the analysis in \S\ref{sec:LSD} for the final EB model (\S\ref{sec:EBmodel}).

Only in this preliminary model, the level of the third light was based on the relative heights of the CCF peaks from \citet{Hebb2010}, and a coeval, lower mass star predicted from stellar models \citep[e.g.,][]{Baraffe2015}. Thus, we utilized a third light that corresponds to  9\%, 11\% and 18\% of the total light for the \vri\ pass bands, respectively, as the dilution in the light curves due to the tertiary. For the final EB model (\S\ref{sec:EBmodel}), we used the levels of third light derived in Sect.~\ref{sec:L3}.

Given the above, we first fitted with PHOEBE \citep{Prsa2005}, the available CTIO light curves to derive the time of mid-transit. We then fitted the two radial velocity curves to derive the EB parameters that are fully defined by the RV curves, namely: $a\sin{i}$, mass ratio $q_{\rm EB}$, and systemic velocity $\gamma_{\rm EB}$ at the time of the UVES observations. These values are reported at the top of  Table~\ref{table:eb}, and remained fixed for the rest of the analysis.  The radial velocity curves and the best-fit RV model are shown in Fig.~\ref{fig:rvs}.
We then reached a solution manually and utilizing the PHOEBE Levenberg–Marquardt (LM) minimization algorithm to determine a model that fits the \vri\ light curves well.  The solution was attained by varying the inclination, the potentials, and the secondary temperature. At each step the limb-darkening coefficients were interpolated for each passband. This provides estimates for the radii and thus surface gravity of the eclipsing components, and the secondary temperature that were used in the determination of the spot properties below (see \S\ref{sec:spots}). Iteratively, such that we derived a consistent solution for the RV curves and the light curves, we also refine the time of mid-transit, which is derived to be $2454972.65085 \pm 0.00016$ days (BJD$_{\rm TDB}$).

\subsubsection{Stellar surface spots in the CTIO 2009 photometry}
\label{sec:spots}

The presence of stellar surface spots is most evident from the out-of-eclipse phases of the light curves.
We measured a peak-to-peak amplitude of 0.14 mag in the U-band, 0.09 mag in the B-band, 0.06 mag in the V-band, 0.04 mag in the \rc-band, and 0.03 mag in the \ic-band. 
These measured amplitudes are at least $5\times$ larger than the
corresponding median photometric uncertainty of the CTIO light curves. 
The amplitude of the variation in the out-of-eclipse light curves of MML~53
increases with decreasing observed wavelength, as expected for stellar surface spots \citep[e.g.,][]{Bouvier1989,GomezMaqueoChew2009}.  Additionally, the eclipsing binary is a detached system, meaning that the components are not interacting (i.e., there is no mass transfer). The eclipsing components 
 are far enough away from each other that reflection effects have 
 amplitudes that are smaller than the photometric precision of each light curve 
\citep[$<$ 5 mmag;][]{Wilson1990},   
and little deformation of the stars occurs ($| r_{equator,i} - r_{pole,i} | < 1\%$).
Finally, the asymmetry of the light curves before and after the secondary eclipse (see Figs.~\ref{fig:lcs} and \ref{fig:ublcs} from phase 0.4 to 0.6) indicates that there is a non-homogeneous distribution of surface brightness in the combined light from the stellar disks (unlike ellipsoidal variation).  The result of these qualitative observations
leads to the conclusion that the observed deviation of the out-of-eclipse phases from the 
relatively flat light curve is most likely due to stellar surface spots.

The depth and shape of the eclipses are affected by the presence of spots, and consequently 
so are the derived radii \citep[e.g.,][]{Covino2004,Morales2010,Windmiller2010} and the temperature ratio \citep{GomezMaqueoChew2009}.  In order
to characterize this in-homogeneity of surface brightness and derive robust 
physical properties, we attempted to model the light curve features with cooler surface spots. 
The spot parameters are degenerate, and we have very little constraint given our data sets on their properties.  We have some information about the longitude of the spots based on the position of the deepest modulation in the out-of-eclipse light curve, and about the temperature of the spots relative to the stellar temperature from the relative depth of the spot modulation as a function of wavelength.  However, the spot size, temperature, and latitude are all highly degenerate and multiple combinations of parameters can easily produce the same light curve variation.    Despite the degeneracy, in order to study the stellar properties, we need to only adopt a single set of parameters that represent the light curve, thus minimizing the effect of the spots on the derived bulk properties of the eclipsing stars (i.e., radius and temperature).

The light curves show two clear regions where independent spots are affecting the light curve.  This prompted us to adopt a two-spot model.  The apparent dip in brightness due the spots is greatest at the primary eclipse causing us to place one spot on the side of the primary star that faces the secondary star (i.e., defined as longitude 0$^\circ$ in PHOEBE).  In addition, to model the region around the secondary eclipse, we placed a second spot on the primary star 135$^\circ$ in longitude from the first spot.  These longitudes remained fixed for the rest of the light curve modeling.  We iterated on the spot positions with the LM solver in PHOEBE and found equally good fits for these positions within $\pm$2$^\circ$ in the longitude, so we adopted these values exactly for our spot longitudes.      

The duration of the spot modulation around the primary eclipse covers a large fraction of the total orbital phase of the light curve.  To fit the large feature that encompasses from about phases $-$0.3 to 0.3 (Figs.~\ref{fig:lcs} and \ref{fig:ublcs}), we could adopt a larger spot at the equator or a smaller spot at a latitude that is closer to the pole.   In order to fit this feature, we needed to choose a relatively large spot at a non-equatorial latitude to create the large duration feature.  As mentioned above, the size of the spot and its latitude are somewhat degenerate, so many combinations of parameters result in equally adequate fits to any given part of the light curve.  Therefore, to model the large feature, we adopted an angular radius of 30$^\circ$ and a latitude of 45$^\circ$. In addition, the light curve is best fitted when the spot is not occulted by the secondary star---again causing us to choose a non-equatorial spot.   

The light curves are well fitted by two spots, both located on the primary stellar surface.  The placement and sizes of the two spots on the stellar surface were optimized to match the shape of the asymmetries in the observed light curves about both eclipses. 
Because the spot temperature is highly degenerate with the spot size \citep[e.g.,][]{GomezMaqueoChew2009},  
the temperature factor (i.e., spot to stellar surface temperature ratio) of each spot was fitted for any given level of third light, because for a fixed size it determines the amplitude of the effect in the light curves due to spots.    
The best-fit spot parameters for the adopted values of third light (see \S\ref{sec:L3}) are given in Table~\ref{table:spots}.  

Other observed evidence that supports magnetic activity and thus the presence of surface spots are: the measured activity indicators in the stellar spectra 
\citep[e.g., H-$\alpha$ is measured in emission;][ and references therein]{Hebb2010},
 the blue-excess in the level of third light \cite[\S\ref{sec:EBmodel}; e.g.,][]{GomezMaqueoChew2012,Gillen2017}, and
 the observed rotational modulation in the WASP light curves.

\begin{table}
\caption{Stellar Surface spot parameters on the primary star}
\label{table:spots}    
\centering                        
\begin{tabular}{c c c c c}       
\hline\hline       \\        
 & Colatitude & Longitude & Radius & Temperature\\ 
 & (deg)	& (deg) & (rad)	& factor \\ 
\hline        \\                
  Spot 1 & 45 & 0 & 30  & 0.94 \\     
  Spot 2 & 90 & 135 & 10    & 0.85 \\  
\hline                                   
\end{tabular}
\end{table}

\subsubsection{Effect of third light on eclipsing component properties}\label{sec:thirdl}

\begin{table}
\caption{Constraint on level of third light from EB light curve model}
\label{table:phoebe3l}    
\centering                       
\begin{tabular}{c c c}
\hline\hline\\
 & Minimum & Maximum \\
 & Dilution & Dilution \\
\hline \\
$L_3/(L_1+L_2+L_3$)\\
in U & 0.0 & 0.55 \\
in B & 0.0 & 0.56 \\
in V & 0.0 & 0.59  \\
in \rc & 0.0 & 0.60 \\
in \ic & 0.0 & 0.63 \\
$i~(^\circ)$ & 81.4 & 90.0 \\
$\chi^{2}_{\rm reduced}$ & 2.2 & 2.5\\
\hline\\
$\Delta$ \mprim$^\dagger$  & \multicolumn{2}{c}{3.3\%} \\
$\Delta$ \msec$^\dagger$ & \multicolumn{2}{c}{3.3\%} \\
$\Delta$ \rprim$^\dagger$  & \multicolumn{2}{c}{1.1\%} \\
$\Delta$ \rsec$^\dagger$  & \multicolumn{2}{c}{1.1\%}\\
$\Delta$ \teffs/\teffp$^\dagger$ & \multicolumn{2}{c}{2.4\%}\\	
\hline
\multicolumn{3}{l}{\tiny{$^\dagger$ Difference in value between models with minimum and maximum dilution}}
\end{tabular}
\end{table}

From the previous analyses, we are certain that there is dilution in the light curves because of the presence of an unresolved third component of MML 53, which at the distance of UCL \citep[140 $\pm$ 2 pc;][]{deZeeuw1999} is a few tens of milli-arcseconds in angular separation from the EB 
\citep[see \S\ref{sec:M3} and][]{Schaefer2018}.
In this section, we explore the effects of the third light level on the physical
properties of the eclipsing components by modeling the eclipsing binary with varying levels of third light.  The largest differences in the EB physical parameters come from the comparison of (a) the case in which there is no dilution in the light curves (i.e., the third light represents 0\% of the total light of the system), and (b) the case in which the inclination of the EB is 90\degree (i.e., the model eclipses are deepest and thus the dilution by the third light has to be the highest to match the observed eclipse depth).  
We modeled these two cases with PHOEBE and show our results in Table~\ref{table:phoebe3l}. The temperature factor of each of the two spots (see \S\ref{sec:spots}) was modified to fit the observed amplitude in the light curves, depending on the level of third light.  
These two extreme cases show that the largest uncertainty in the masses of the eclipsing components due to the level of third light is $\sim$3\%. In the case of the radii, this uncertainty is $\sim$1\%. We also find that in the case of maximum dilution, the level of third light required to match the observed eclipse depths does not significantly decreases toward the bluer bands as would be expected for a lower-mass tertiary, as is suggested by the height of the peaks of the CCF of the combined spectra and as is determined in \S\ref{sec:M3}.  In fact, the third light level required is relatively flat in all passbands, which indicates an excess in the level of third light in the bluer bands in the case of a lower-mass tertiary.

We consider the uncertainty on the physical properties of the eclipsing stars 
due to the amount of third light to be much smaller than these values because: (a) the fit to the observed light curves at the two extremes is worse than our best fit model (best $\chi^{2}_{\rm reduced} \approx$1.5 and Table~\ref{table:phoebe3l}); (b) both cases are not physical, because we know that the tertiary  exists and it is diluting the light curves, and it is a lower-mass star gravitationally bound to the system; and (c) we do have constraints on the kind of star that is diluting the light curves (see \S\ref{sec:M3}), even if the amount of light we adopt is model dependent.

\subsection{Binary-tertiary model} 
\label{sec:M3}

    \begin{figure}
    \centering
    \includegraphics[width=\hsize]{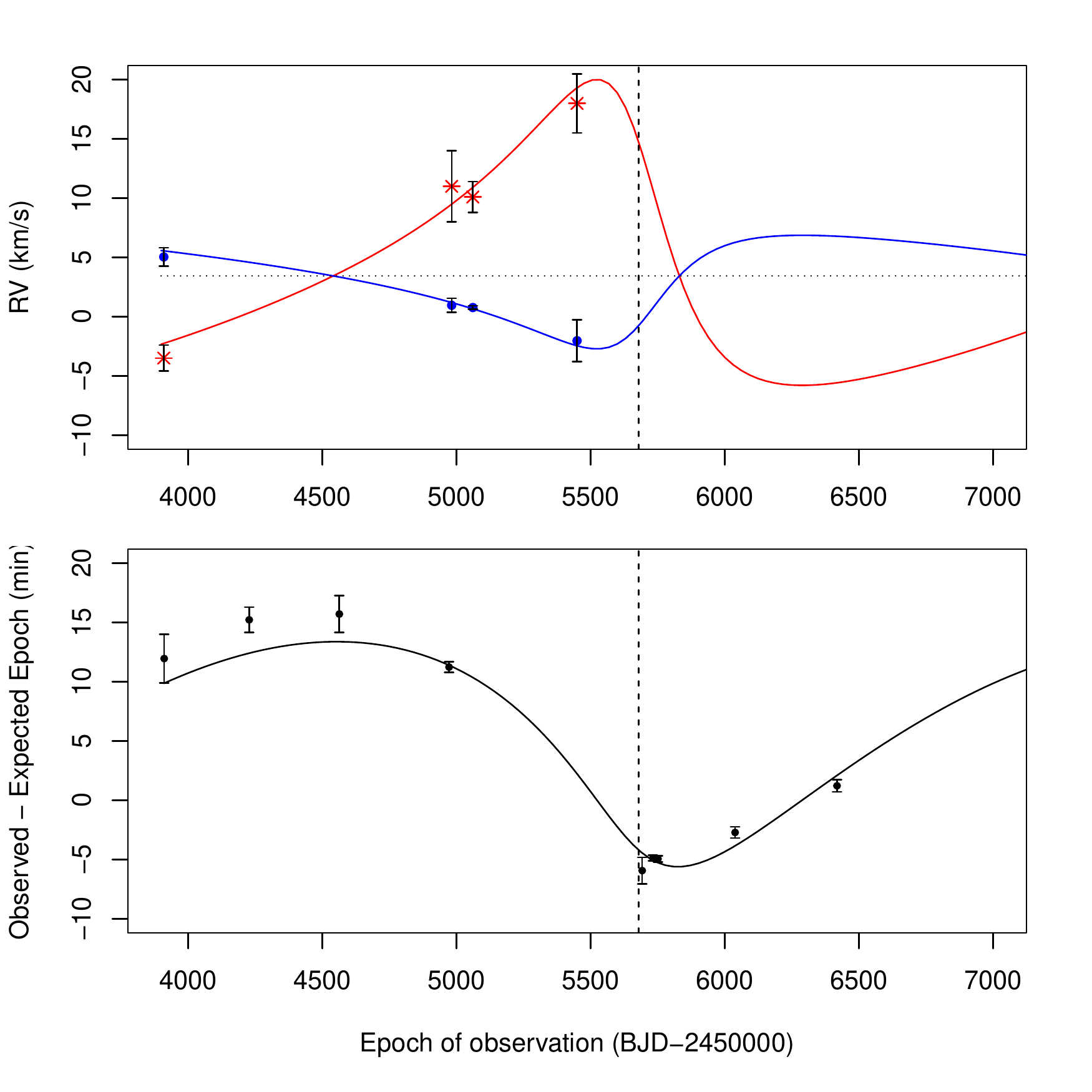}
    \caption{{\bf Top}: Radial velocities for eclipsing binary system and tertiary star.  We present the observed radial velocity curve for the tertiary star (red asterisks) and the systemic radial velocity of the eclipsing binary (blue circles) compared to the best fitting model radial velocity curves derived from the MCMC analysis described in Sect.~\ref{sec:M3}.  
    {\bf Bottom}: Measured eclipse timings offsets.
    We show the measured change in the time at which the eclipse occurred due to the light-time travel effect as it orbits
    the center of mass of the three-body system compared 
    to the best-fit Keplerian model (solid black line). The vertical dashed line represents the time of periastron passage.  This does not occur at the maximum radial velocity because it occurs at the maximum 3-dimensional velocity which is close to, but not exactly the same as the line-of-sight peak.} 
      \label{fig:m3model}%
     \end{figure}

    \begin{figure}
    \centering
    \includegraphics[width=\hsize]{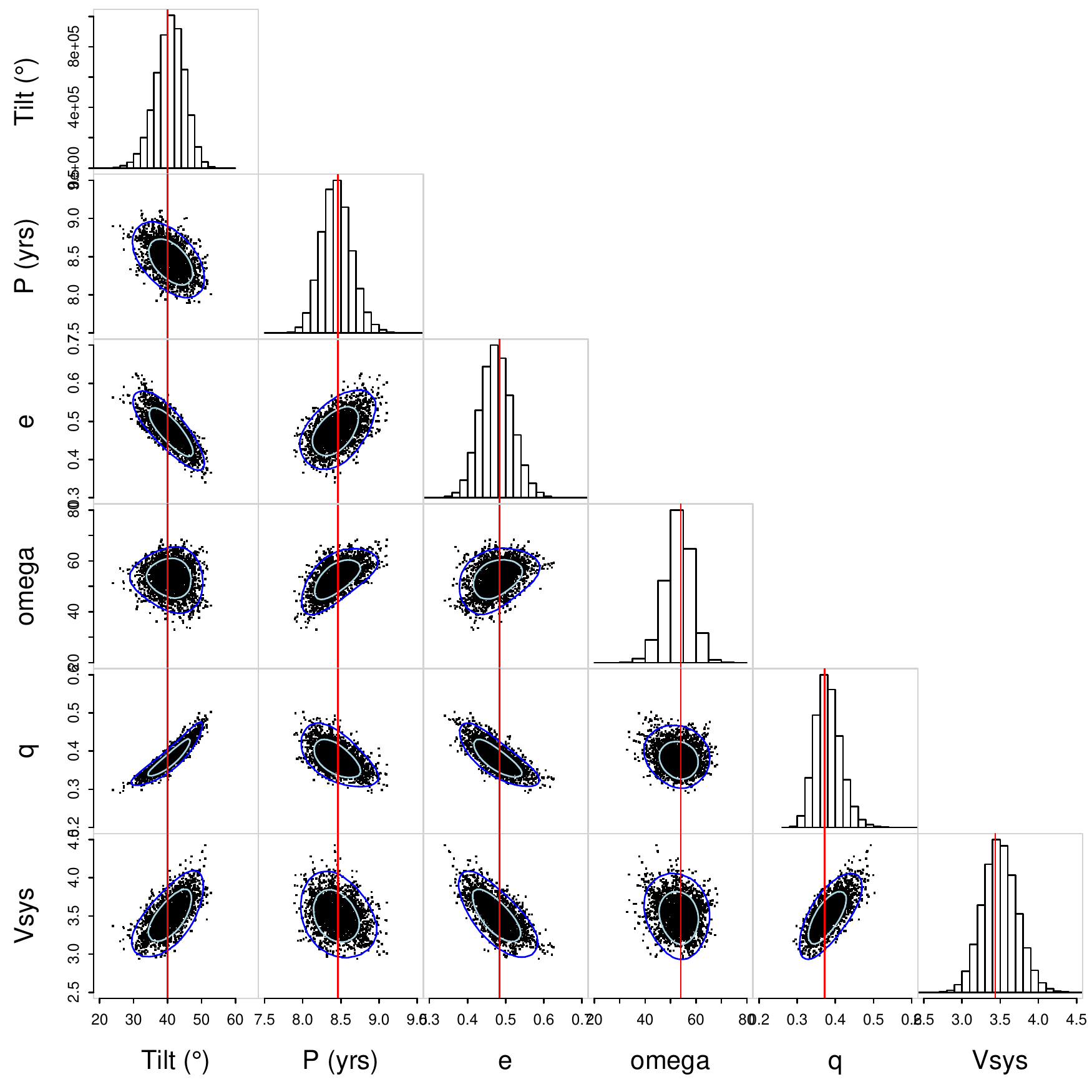}
    \caption{Final MCMC distributions for EB--tertiary orbit. The Keplerian model that best fits the RV measurements and the eclipse timings is determined from this analysis. The parameters that provide the lowest $\chi^2$ solution are marked with the red vertical lines and their uncertainties are given by each corresponding distribution. }
      \label{fig:m3mcmcpanel}%
     \end{figure}

The tertiary is visible in the spectrum as an independent component, but variations in its radial velocity compared to the systemic radial velocity of the binary confirm that the tertiary is part of a triple system.  In addition, the timing of the eclipses of the binary components vary periodically due to the binary's orbital motion around the tertiary star \citep{Hebb2010}.  
We used the all of the available light curves to investigate eclipse timing variations caused by the wide tertiary star.

\subsubsection{Eclipse timings}
\label{sec:epochtimes}

The rectified light curves derived from the 2006-2013 WASP data and the two individual eclipse events obtained with the FTS telescope in 2011 are used to measure variations in the time of minimum light for the MML~53 EB due to its motion around a common center of mass with the tertiary companion. 

\begin{table}[h]
\caption{Times of light curve minima for primary eclipse}\label{table:timings} 
\begin{tabular}{l c}
\hline\\
Dataset	& Time of minimum  \\
 & (BJD$_{\rm TDB} - 2\,450\,000$)\\
\hline\hline\\
WASP 2006     & 3911.1182 $\pm$ 0.0007 \\   
WASP 2007     & 4227.9021 $\pm$ 0.0004 \\                   
WASP 2008     & 4563.5651 $\pm$ 0.0005 \\
CTIO May 2009  & 4972.6509 $\pm$ 0.0002 \\        
WASP 2011     & 5692.2197 $\pm$ 0.0004 \\   
FTS-20110619   & 5732.0764 $\pm$ 0.0001 \\                         
FTS-20110709   & 5750.9574 $\pm$ 0.0001 \\                               
WASP Feb 2012 & 6038.3701 $\pm$ 0.0002  \\                       
WASP 2013     & 6418.0912 $\pm$ 0.0002  \\
\hline
\end{tabular}
\end{table}

We fitted the rectified light curves from individual seasons of data (including all observed primary and secondary eclipses) using the fast, analytic EB modeling code EBOP \citep{popperetzel,southworth2007}. This program treats the stars as detached, nearly spherical geometric shapes in order to derive the orbital parameters (i.e., period, epoch, eccentricity) of the binary and some eclipse parameters that are directly related to the shape of the light curve (i.e., sum of the stellar radii, surface brightness ratio), but it does not provide direct physical properties of the stars (i.e., individual temperatures and stellar radii).  Since we require only the time of minimum light and not a full EB model solution from these data, this program is suitable for the analysis.     

While analyzing the light curves with EBOP, we used the Levenberg-Marquart fitting option and allowed the time of minimum light to be a free parameter, but the orbital period of the binary, the stellar masses, the eccentricity, and the secondary parameters (limb darkening, gravity brightening, reflection coefficients, and third light contribution) remained fixed to the values derived in the final EB model from PHOEBE.  
The relative sum of the stellar radii, the surface brightness ratio, and the inclination angle were allowed to vary in each case in order to provide sufficient freedom to find the best fitting model light curve while accounting for small variations in the relative eclipse depths due to the filter, the contribution of third light to that filter, and starspots. 

The best fitting model light curves are overplotted in red on the phase-folded, rectified light curves shown in Figs.~\ref{fig:swaspphot} and \ref{fig:ftsphot}, and the final minimum eclipse times derived from these fits are
reported in Table~\ref{table:timings}.  The 2013 WASP data consists of three independent light curves from different WASP cameras that cover the same time period.  Each light curve was fitted independently with EBOP, and the weighted average of the three eclipse time measurements is reported in the table.  The uncertainty on the 2013 measurement is the standard error of the mean of these three values, but all other reported uncertainties come directly from of the final EBOP results.  All final epoch times are converted to BJD$_{\rm TDB}$ time units using the routines provided online by Jason Eastman \citep{BJDTDB}.  We note that the WASP data obtained on this target in the summer of 2012 is not shown in the figure or used in the analysis because no primary or secondary eclipse minima were observed during that time period.  

\subsubsection{Orbital solution of binary-tertiary system}

We combine the eclipsing timing epochs in Table~\ref{table:timings}  with the radial velocity measurements described in \S\ref{sec:LSD} to constrain the orbital parameters of the binary-tertiary system. We also incorporate into the fits the recent measurements described in \citet{Schaefer2018} of the angular separation of the binary and tertiary along with the angular position change of the components in the plane of the sky.  To do this, we developed a program that solves for the orbital parameters of a two-body Keplerian system by treating the MML~53 eclipsing binary as a single mass, $M_B = M_1+M_2$, in orbit with the tertiary star, $M_3$, around a common center of mass.     This is justified since the separation between the binary components is less than 1\% of the separation between the binary and the tertiary.  Incorporating the effect of both binary components on the motion of the tertiary is needlessly complicated given the final uncertainties on the orbital parameters of greatest interest.       

There are seven independent parameters that are used to define the orbital motion the binary-tertiary system:  the orbital period, $P_3$; the mass ratio, $q_3 = M_3/(M_B$); the eccentricity, $e_3$; the argument of periastron, $\omega_3$; the tilt of the orbital plane from the observer's line-of-sight, $\theta_3$; the systemic velocity of the triple system, $\gamma$; and the time of periastron, $t_{\rm peri}$.   In addition, we assume the mass of the binary, $M_B = 1.9307$ \msun\ 
is known, and we used Kepler's Law to derive the orbital separation, $a_3$, between the binary and the tertiary components.  These parameters were used in the model generating engine of our program to produce synthetic three-dimensional velocities and positions as a function of time for the binary and tertiary components.   In order to identify the optimum values of these parameters that best reproduce the observations, the model generating engine was wrapped by an affine invariant Markov chain Monte Carlo (MCMC) sampler that was integrated into the program itself \citep{ForemanMackey2013}. 

The program allows the MCMC algorithm to explore the parameter space from a random starting point for all parameters.   For each MCMC trial, it uses the known $M_B$ along with the adopted $P_3$, $e_3$, $q_3$, and $t_{\rm peri}$ values for that trial to find the true anomaly of the binary and tertiary components in their eccentric orbit.  The true anomaly combined with the $\gamma$ velocity, and the orbital separation, $a_3$, were used to determine the positions and velocities of the binary and tertiary masses in the ellipse reference frame (with the orbital angular momentum axis pointed perpendicular to the observers line of sight). The ellipse frame was then rotated by $\omega_3$ and tilted down from the z-axis by $\theta_3$ to achieve the final three-dimensional positions and velocities of the components at each time of an observation. The $\omega_3$ was measured from the center of mass to the tertiary star orbit, as for visual binaries.     To calculate the $\chi^2$ value of that trial set of parameters, the program incorporates a comparison between the synthetic line-of-sight velocity of both components to the observed radial velocities at each time.  It also uses the line-of-sight position of the binary component relative to the center of mass divided by the speed of light to compare to the observed eclipse timing offsets.  Furthermore, the distance between the EB and tertiary in the plane of the sky was compared to the measured angular separation multiplied by the Gaia (data release 2) distance of 130.2~pc and added to the $\chi^2$ value.  Lastly, to incorporate the angular motion of the system around the common center of mass in the plane of the sky, which is reported in \citet{Schaefer2018}, we first solved for the angular offset which minimizes the difference between the measured angles and those in the model.  This is necessary because the orientation of the model angles must match the reference point defined in the observations.  This angular offset corresponds to the longitude of the ascending node, $\Omega_{\rm ascending}$, which is the angle from the reference direction north to the line connecting the center of mass and the orbital plane of the tertiary component when it crosses the plane of the sky in the direction away from the observer.  This angular offset was then applied to the model angles in the plane of the sky before comparing them to the measurements. For completeness, we also report the angular semi-major axis between the visual components, $a_{\rm angular}$.  According to our model, the maximum separation occurs close to the first \citet{Schaefer2018} measurement obtained in 2014.    

We ran the model generating engine in the affine invariant MCMC sampler with 1000 walkers.  In applying this algorithm, we took special care with certain parameters.  Due to the degeneracies in the system, we only allowed $\theta_3$ to vary from $0 - \pi/2$.  We adopt $\sqrt{e_3} \cos{\omega_3}$ and $\sqrt{e_3} \sin{\omega_3}$ as sampling parameters and converted these values to $e_3$ and $\omega_3$ in order to avoid the Lucy-Sweeney bias \citep{Lucy1971,Eastman2013}.  
The mass ratio, $q_3$ can vary outside of the range from $0.0 - 1.0$ when updating its value at each step, so we rejected all values greater than 1.0 after the MCMC is complete since these values are unphysical in our model and in our understanding of the MML~53 system.  

We allowed each walker to run for 30,000 trial steps of which 30-33\% were accepted resulting in a final distribution of 9000-10,000 accepted steps per walker.  After examining the output MCMC file of accepted parameters, we chose to remove the first 300 accepted steps from each walker as it constitutes the burn-in phase.  Furthermore, only 836 of the initial 1000 walkers converge to a single solution at the global minimum of the $\chi^2$ space.  The remaining walkers appear to get stuck in local minima at much higher $\chi^2$ values with much lower acceptance rates.  We removed these walkers from the final distribution, but considered the remaining walkers to be converged (shown in Fig.~\ref{fig:m3mcmcpanel}).  The final best fitting model has a $\chi^2_{min} = 26.88$. The best fitting parameters and their $1\sigma$ uncertainties are shown in Table~\ref{tab:m3params}.  The mass of the tertiary, $M_3 = 0.72^{+0.16}_{-0.09}$~\msun, is derived by multiplying $M_B = 1.9307$~\msun and the newly derived mass ratio $q_3 = 0.37^{+0.08}_{-0.05}$.   In Fig.~\ref{fig:m3model}, we show the best fitting model radial velocity curves of the EB and third star system, and eclipse timing curves compared to the observations.

    \begin{table}
       \caption[]{Orbital solution of EB and tertiary}
          \label{tab:m3params}  %
          \vspace{-0.75cm}
          $$ 
          \begin{array}{p{0.5\linewidth}lll}
             \hline\hline
             \noalign{\smallskip}
             Fitted Parameter      &  Value  & +1 \sigma & -1 \sigma \\
             \noalign{\smallskip}
             \hline
             \noalign{\smallskip}
             $P_3$ (years) & 8.5 & 0.4  & -0.4             \\
             $q_3$ & 0.37  & 0.08  & -0.05             \\
             $e_3$ & 0.48  & 0.08  & -0.09             \\
             $a_{\rm angular}$ (mas) & 56.2 & 5 & 5 \\
             $\omega_3$ ($^\circ$)& 54 & 9  & -12      \\        
             $\theta_3$ ($^\circ$)& 40 & 9  & -8      \\        
             $\gamma$ (km/s) & 3.4 & 0.6  & -0.5      \\   
             $\Omega_{\rm ascending}$ ($^\circ$) & 174 & 12 & -12  \\
             $t_{\rm peri} (BJD_{\rm TDB}-2\,450\,000) $ & 5679.64 & 0.4 & -0.4 \\
             \noalign{\smallskip}
             \hline
          \end{array}
      $$ 
    \end{table}

\subsection{Third light determination} \label{sec:L3}

Light from the unresolved tertiary star causes both the primary and secondary eclipses of the EB to appear shallower than they should.  This directly affects the derived orbital inclination angle of the EB, which indirectly influences the individual masses and radii.    Thus, in order to derive accurate fundamental properties of the primary and secondary components from the eclipsing light curve, 
quantitative values of the light contributed by the tertitary star in each filter must be incorporated into the EB model.  The third light values applied here were derived from the latest theoretical stellar evolution models from \citet{Baraffe2015} based on the mass and age of the tertiary star.   These isochrones provide \vri\ band absolute magnitudes as a function of mass and age that are interpolated to find the flux contribution of the third star relative to primary and secondary components in each filter.  

An initial guess for the third light was derived from the relative height of the cross-correlation peaks in the 2006 FEROS spectrum.  This third light value is used to perform a preliminary EB model which results in masses and luminosities for the primary and secondary components as described in \S\ref{sec:pre-eb}.  This preliminary value of third light does not affect the final values of third light used nor the EB physical properties. To derive the third light used in the final EB model, the primary and secondary star masses were combined with the binary-tertiary mass ratio, $q_3$, to find the mass of the tertiary star. With preliminary masses for all three components known, we implemented a quadratic interpolation of the published isochrones that are less than $50$~Myrs old at each of the component masses to derive a series of theoretical \vri\-band absolute magnitude values for each star as a function of age.  

At this point, we could interpolate the mass tracks
at the independent age measured for the Upper Centaurus Lupus cluster ($\sim 16$~Myr) to find the relative luminosities of the stellar components, however we chose instead to find self-consistent values that match both the luminosities derived from the EB model and a single theoretical isochrone for all three stars.   The EB model provides a measured value for the primary-to-secondary flux ratio in each filter based on the temperatures and radii of each star.  We used this value as a constraint and convert the theoretical absolute magnitude values into an array of primary to secondary star flux ratios ($L_2/L_1$) in the \vri\-bands  as a function of age.  We then interpolated the models at the measured $L_2/L_1$ value from the EB model in each filter. 

This provides an independent age estimate for the system, which is the same for all three filters within 1~Myr.  Finally, we interpolated the \vri\-band absolute magnitude values for each component at that age and calculate the tertiary star's relative contribution to the overall light of the system in each filter.  

This is an iterative process in which the EB model, the binary-tertiary model, and the third light calculation are performed in consecutive order until the masses and relative light contributions of the three components have converged within the uncertainties.  The relative \vri\ band fluxes determined for each star after several iterations are shown below including the third light contribution ($L_3/(L_1+L_2+L_3$)) that is used in the final EB model.  As a consistency check, we measured the relative surface brightness from the CCF of the FEROS spectrum fitting a three-Gaussian model obtaining 0.59:0.23:0.18, which are fully consistent with the values presented in Table~\ref{table:thirdlight}.

\begin{table}
\caption{Flux of each stellar component relative to the total flux}
\label{table:thirdlight}    
\centering                       
\begin{tabular}{l c c c}
\hline\hline\\
     & V-band & R-band & I-band \\
\hline \\
$L_1/(L_1+L_2+L_3$)  & 0.65  & 0.61 & 0.55\\
$L_2/(L_1+L_2+L_3$)  & 0.24  & 0.26 & 0.28 \\
$L_3/(L_1+L_2+L_3$)  & 0.11 & 0.13 & 0.17 \\
\hline
\end{tabular}
\end{table}

  \begin{figure}[!h]
   \centering
   \includegraphics[width=\hsize]{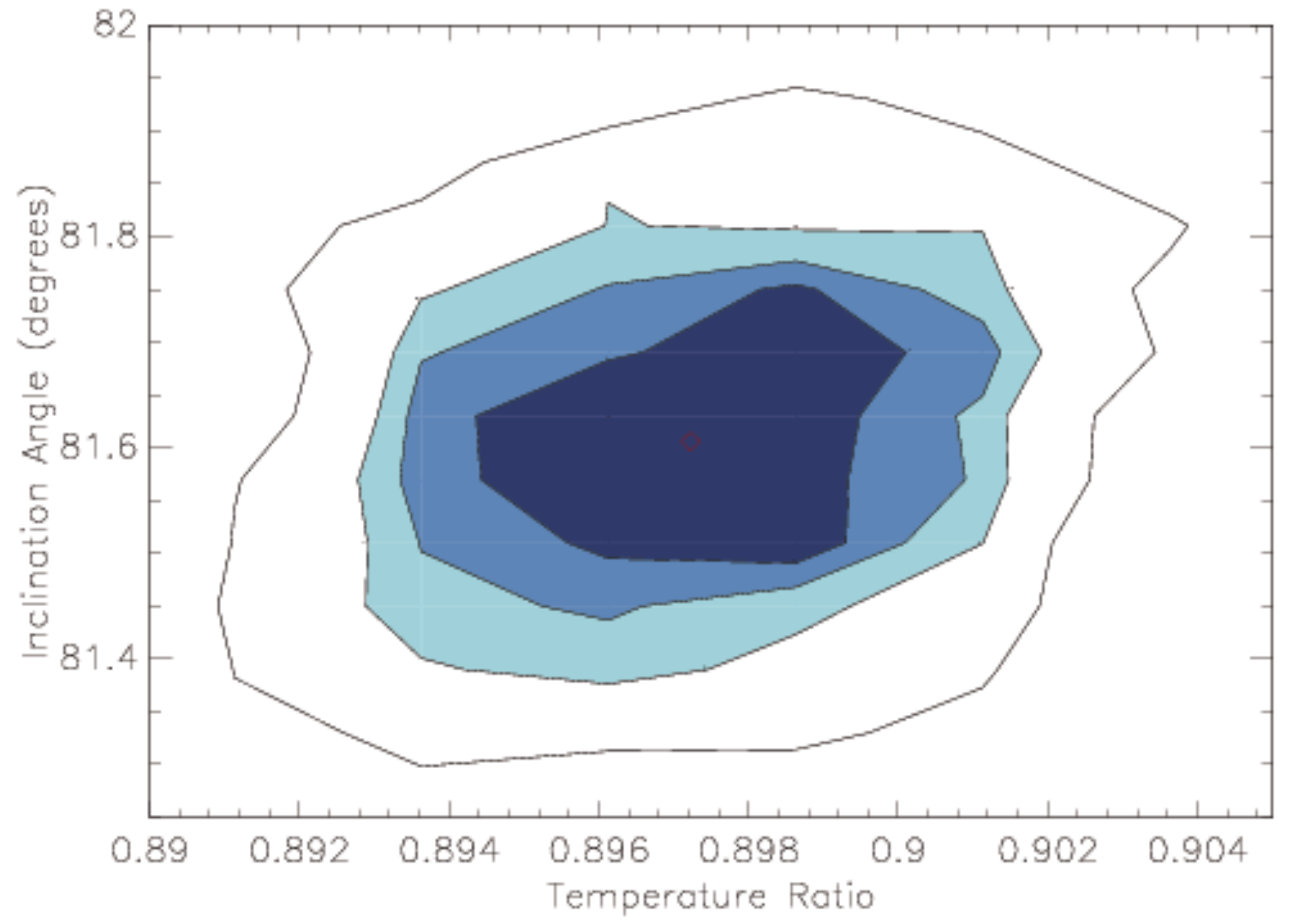}
   \includegraphics[width=\hsize]{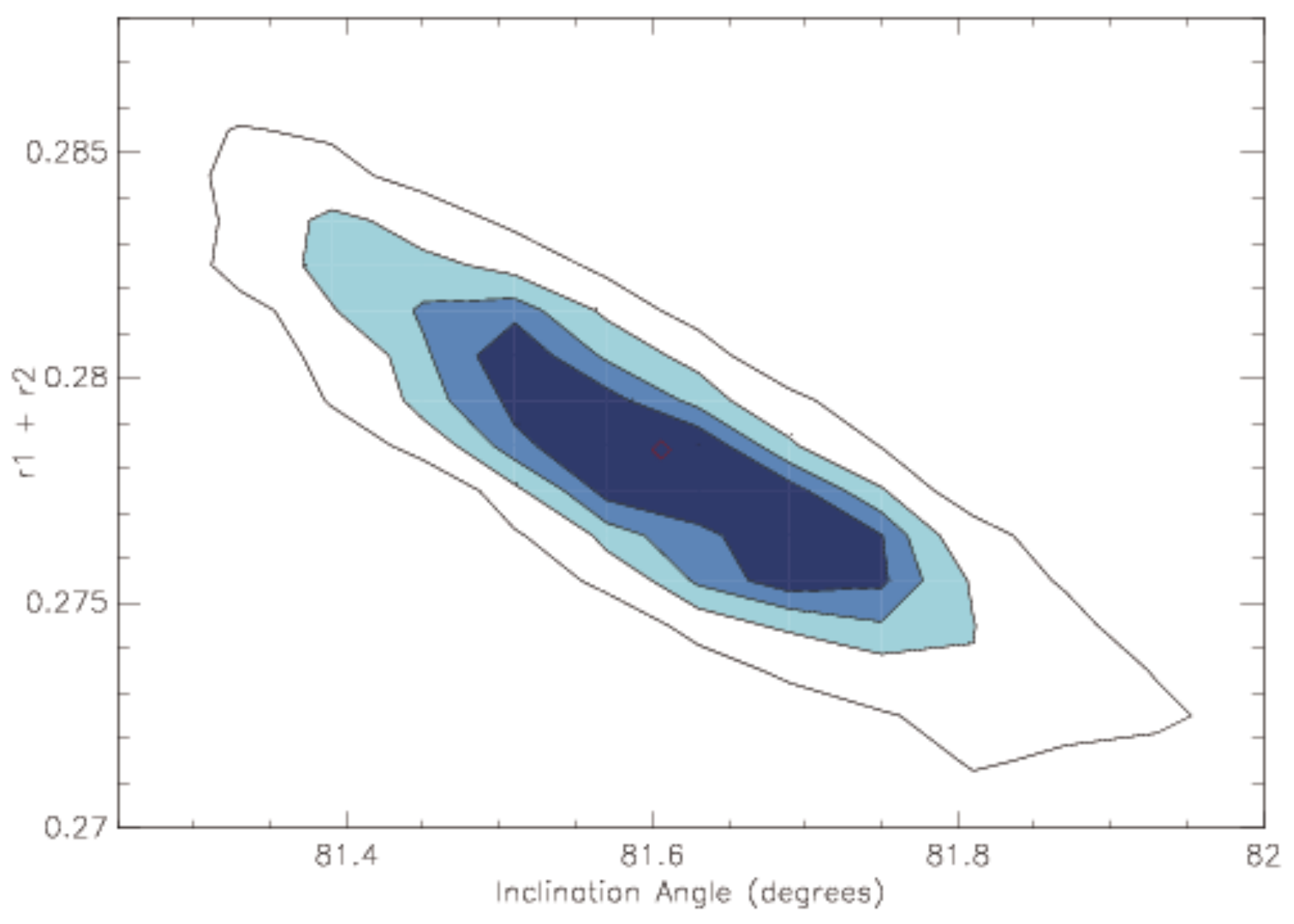}
      \caption{Confidence levels for inclination, Sum of the fractional radii and temperature ratio derived from the \vri\ light curves. {\bf Top}: We show from the center the 1--, 2--, 3-- and 5--$\sigma$ contours of the  inclination of the eclipsing binary orbit and the temperature ratio (\teffs/\teffp).  The confidence levels were determined from the exploration of the $\chi^2$--space, and the best-fit solution to the RV and \vri\ light curves is the one with the lowest $\chi^2$ (marked by the red diamond in both panels).  
 {\bf Bottom}: We show the sum of the fractional radii (i.e., $r_1 + r_2$) as a function of the inclination angle. The contours represent the same confidence levels  as in the top panel. }
         \label{fig:contours}
   \end{figure}

\subsection{From EB model: masses, inclination, sum of the radii and temperature ratio}
\label{sec:EBmodel}

Utilizing the eclipsing binary tool PHOEBE \citep{Prsa2005}, we modeled both the UVES radial velocity measurements for the primary and secondary components, and the CTIO light curves.  We only fitted the \vri\ light curves to derive the physical properties of the eclipsing components to limit the uncertainty introduced by the level of third light (\ref{sec:L3}). The \vri\ light curves are shown in Fig.~\ref{fig:lcs} and compared to the best-fit model described in this section. 
Not only are the UB bands not included in the evolutionary models that determine the amount of expected dilution due to the tertiary, but we find that the U-band has an additional third light contribution than would be expected for a less massive star. Thus, the UB bands were not used to determine the best-fit model, but are shown for reference in Fig.~\ref{fig:ublcs}, and are included in this paper to distribute the full CTIO photometric dataset to the community.

Adopting the spot sizes and placements from \S\ref{sec:spots}, the level of third light from \S\ref{sec:L3}, and the \teffp\ (= 4880~K) from the spectral 
disentangling (\S\ref{sec:LSD}), 
we randomly sampled 80,000 times the following parameters (and ranges):
orbital inclination (79.5 $\leq$ $i$ $\leq$ 84.1\degree);
the primary potential (5.9 $\leq$ $\Omega_1$ $\leq$ 10.0);
the secondary potential (5.7 $\leq$ $\Omega_2$ $\leq$ 9.7), 
and the secondary temperature (4245 $\leq$ \teffs $\leq$ 4545~K). 
For each combination of these parameters, 
we fitted the temperature factor of each stellar spot and the overall luminosity for each light curve, interpolated the limb-darkening coefficients for each band, and calculated the $\chi^2$ of each model.

From the resulting multidimensional $\chi^2$-space and considering the detached and circular orbit of the eclipsing binary, we obtained confidence levels (shown in Fig.~\ref{fig:contours}) for the properties that are derived directly from the light curves, namely: the inclination angle, 
$i$; the temperature ratio, \teffs/\teffp from the relative depth of the eclipses, and the sum of the fractional radii, $r_1 + r_2$ = (\rprim+\rsec)/a  from the duration of the eclipses. 
In the case of MML~53 because the eclipsing binary orbit is circular and the eclipses are V-shaped,  we are able to constrain the sum of the fractional radii from the light curves and not the radius ratio. Given this degeneracy, although we sampled the primary and secondary potentials for the parameter-space exploration, we do not report the individual values as they are not significant, as is well known for EBs in circular orbits \citep{Kallrath2009}.
The best-fit model to the \vri\ light curves (see Fig.~\ref{fig:lcs}) was identified for having the lowest $\chi^2$, with a corresponding reduced-$\chi^2  \approx$ 1.5. 

In the case of the U and B light curve models, we adopted the \vri\ best-fit solution and 
 fitted the level of third light to match the amplitude of the variation due to spots and the depth of the eclipses. 
We find that the B-band is well fitted with a dilution by the tertiary of 
8\% of the total luminosity of the three-body system, while the U-band requires a 15\% dilution. 
 Given that the tertiary is less massive (and thus, redder) than the eclipsing components, we would expect the level of dilution due to the tertiary to decrease toward bluer wavelengths.  An additional blue component in the third light levels could be due to accretion on to the tertiary, as has been identified in at least another PMS eclipsing binary Par~1802 \citep{GomezMaqueoChew2012}. Similarly, a  $u$-band excess has been shown by the eclipsing components of CoRoT 223992193 \citep{Gillen2017}. Other young, single stars have been observed to have UV excess and optical spectra accretion features \citep[e.g.,][]{Findeisen2010}. 
Figure~\ref{fig:ublcs} shows the model and observed light curves in the U and B passbands. We have not included the residuals to the models because these light curves are not used to derive the best-fit model.

\subsection{Derivation of the semi-major axis, the individual radii and the secondary temperature}\label{sec:radii}

Based on the resulting parameters and their associated uncertainties from the EB model to both the RVs and \vri\ light curves (\S\ref{sec:EBmodel}), 
we calculated the physical properties of interest, namely the semi-major axis, the primary and secondary radius, and the effective temperature of the secondary component. All values are summarised in Table~\ref{table:eb}. 

To determine the physical scale of the orbit, we utilized the parameters derived directly from the RV curves ($a\sin{i}$ and $q$) and their formal uncertainties to the fit together with the values and uncertainties from the confidence levels of the quantities that depend solely from the light curves to derive the physical properties of the eclipsing components, their orbit, and their corresponding uncertainties. Specifically, we derived the semi-major axis of the eclipsing orbit from the definition of $a\sin{i}$ and the measured $i$. Once $a$ was determined, we derived the individual masses  and the total mass from $q$, orbital period and $a$ through the equations of Keplerian motion. 

Given the circular orbit of the MML~53 EB, the grazing nature of its eclipses and the contamination of the photometry by the third star, we require an external constraint in order to derive the individual radii of the eclipsing components \citep[e.g.,][]{Kopal1959,Stassun2004,Stassun2014}. However, in the case of MML~53, the flux ratio derived from the CCF is uncertain, and thus, it was not utilized in this analysis to derive individual radii as the external constraint.   
The primary radius was instead determined from the measurement of the \vsini\ of the primary component from the LSD analysis of the high-resolution UVES spectra (\S\ref{sec:LSD}), the inclination $i$ from the EB model (\S\ref{sec:EBmodel}), 
and that the primary star is synchronized and its  spin-axis is aligned with its orbital motion (\S\ref{sec:pre-eb}), 
given that by definition \vsini$_1$ = $2\pi~\rprim~\sin{i}/P_{\rm rot,1}$.  

The secondary radius was derived from the sum of the fractional radii (Fig.~\ref{fig:contours}, bottom panel), $a$, and the primary radius. 
With the individual radii and masses of the eclipsing components, the surface gravities for the eclipsing components (\loggp\ and \loggs) were derived readily utilizing the fundamental constants from \citet{Prsa2016}. 
As a consistency check, we calculated the secondary radius from the LSD measurement of its \vsini\ and find it to be in agreement within the uncertainties with the secondary radius reported in Table~\ref{table:eb}.

The secondary temperature was derived from the spectroscopically-determined primary temperature \teffp\ (\S\ref{sec:LSD}), and the temperature ratio resulting from the EB model (Fig.~\ref{fig:contours}, top panel). We present our measurements for the physical properties of the eclipsing stars derived from the best-fit eclipsing binary model, adjusting both the RVs and \vri\ light curves in Table~\ref{table:eb}.

\begin{table*}
\caption{Physical properties of MML53 eclipsing stars and their orbit}             
\label{table:eb}      
\centering          
\begin{tabular}{l c c c }     
\hline\hline       
\multicolumn{2}{c}{Parameter} & Value & Units\\ 
\hline                    
Orbital period & \porb$^\ddagger$ &  2.097892  $\pm$  0.000005 & days\\ 
Eccentricity & $e$ & 0. &(fixed)  \\  
Mass ratio & $q_{\rm EB}$ = \msec/\mprim & 0.8565  $\pm$  0.0034  & \\
Systemic velocity & $\gamma_{\rm EB}$ $^\dagger$ & 0.77  $\pm$  0.15 & \kms \\
Semi-major axis & $a\sin{i}$ & 8.492  $\pm$  0.017 & \rsun \\
& $a$ & 8.584  $\pm$  0.018 & \rsun\\
 & $a$ & 0.03992  $\pm$  0.00008 & au \\  
Binary total mass & $M_B$ & 1.9307 $\pm$ 0.0119 & \msun\\
Inclination & $i$ & 81.61  $\pm$  0.12 & \degree \\
Sum of fractional radii & $r_1 + r_2$ & 0.2784  $\pm$  0.0027 & \\
Temperature ratio & \teffs/\teffp & 0.8972 $\pm$  0.0018 & \\
Primary mass & \mprim & 1.0400  $\pm$  0.0067 & \msun \\
Primary radius & \rprim &  1.283  $\pm$  0.043 & \rsun \\ 
Primary surface gravity & \loggp &   4.24  $\pm$  0.03 & dex (cgs) \\ %
Primary temperature & \teffp & 4880 $\pm$ 100 & K\\ 
Secondary mass & \msec & 0.8907  $\pm$  0.0058  & \msun \\
Secondary radius & \rsec & 1.107  $\pm$  0.049  & \rsun \\ 
Secondary surface gravity & \loggs &  4.30  $\pm$  0.04 & dex (cgs) \\
Secondary temperature & \teffs & 4379  $\pm$  100 & K \\
\hline       
\multicolumn{4}{l}{$^\ddagger$ The orbital period was adopted from the analysis of \citet{Hebb2010}.}\\
\multicolumn{4}{l}{$^\dagger$ Systemic velocity of the EB components at the time of the UVES RV observations.}\\
\end{tabular}

\end{table*}

\section{Summary and discussion} \label{sec:discussion}

\begin{figure}
\centering
\includegraphics[width=1.0\hsize]{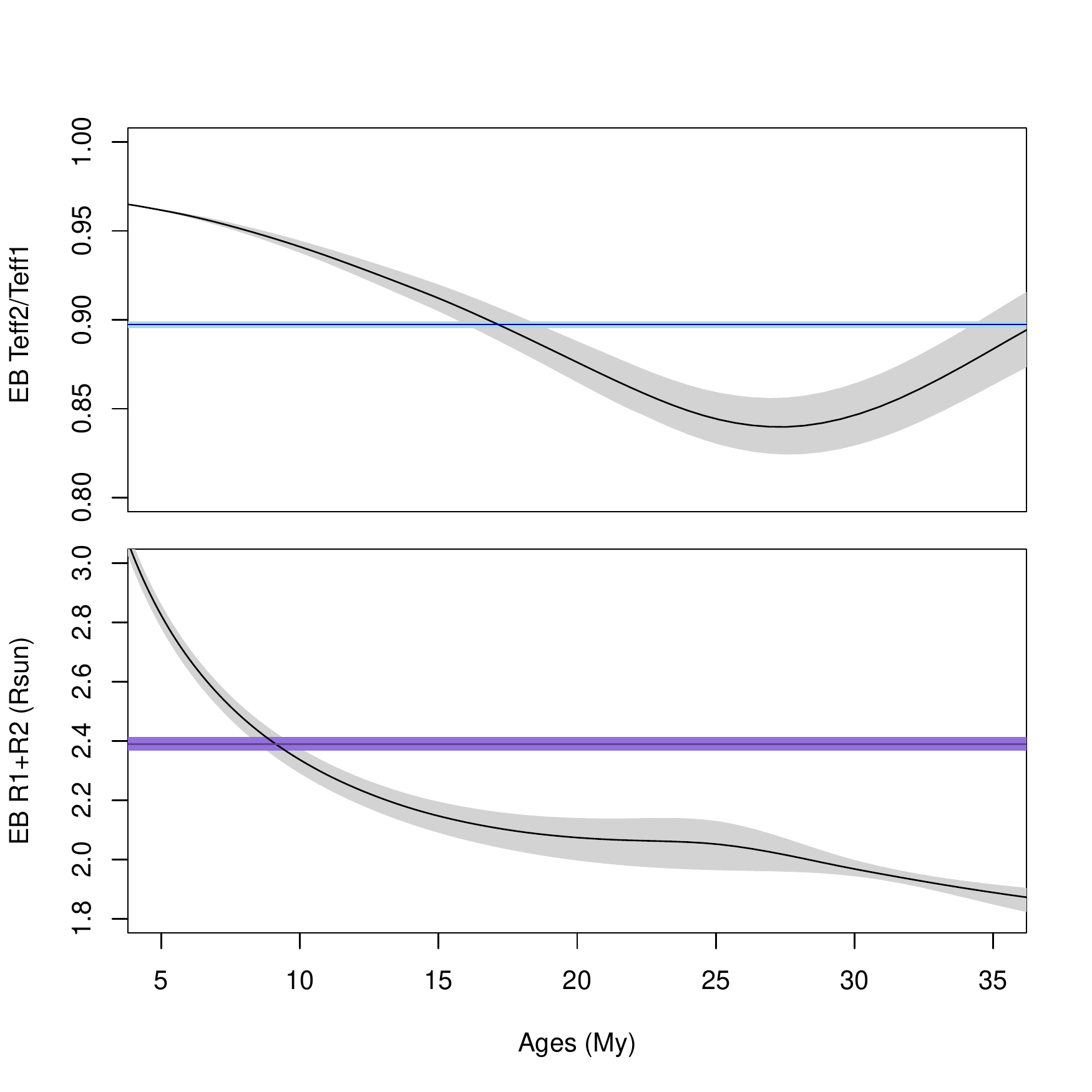}
\caption{We show the constraints on the age of the MML 53 system given by the comparison of the direct measurements of the sum of the radii and the temperature ratio of the eclipsing components against the theoretical evolutionary models of \citet{Baraffe2015}.  The models were interpolated at the measured masses of the eclipsing component stars, \mprim\ and \msec.  The top panel shows the value of the temperature ratio for stars of these two masses as a function of age (black line).  The gray region includes the range of model values allowed for the 1-$\sigma$ uncertainties on the masses.  The horizontal, blue shaded area is the measured temperature ratio and its uncertainty derived from the EB model.
Similarly to above, the bottom panel shows in the black solid line the evolution of the sum of the radii (1-$\sigma$ uncertainty in gray area), as compared to the direct measurement of the sum of the radii and its uncertainty in purple. }
\label{fig:age}
\end{figure}

MML 53 is a gravitationally bound hierarchical triple system where all three components are in the pre-main sequence.  It consists of a close eclipsing binary 
composed of a 1.0400~\msun\ primary star 
and a $0.8907$~\msun\ secondary star in a 
$P_{\rm EB} \sim2.09$~day orbit, and a 
distant, lower mass ($\sim$0.7~\msun) star in a longer period  ($\sim$8.5 yr)   orbit. 
The masses of the eclipsing components have been determined with $<$1\% precision, which allows the mass of the tertiary to be measured to $\sim$20\%. 
Additionally, our analysis of the EB allows us to measure the radii of its components to be 1.237 and 1.153~\rsun\ for the primary and secondary stars with a precision of 2.7\% and 3.5\%, respectively.
We also measure the individual temperature of the primary from the spectral analysis to be 4880 $\pm$ 100~K (2\% precision), and derived from the temperature ratio that of the secondary star to be 4380 $\pm$ 100~K (2\% precision). Although MML 53 shows all the axes of complexity for pre-main sequence EBs (higher multiplicity, spots, possible accretion), we are able to measure precisely the individual properties of the eclipsing stars and constrain for the first time the mass of the tertiary.

 Furthermore, our analysis permits the even more precise, direct measurement of two EB quantities: (1) the sum of the fractional radii (0.9\% precision) from the duration of the eclipses, and (2) the temperature ratio (0.2\% precision) from the relative depth of the eclipses in each passband. These two quantities derived from our eclipsing binary model are robust measurements, and are independent from the light contamination by the third star. Thus, the comparison of these two direct measurements to those predicted by theoretical evolutionary models is important. 
 Their evolution predicted by the \citet{Baraffe2015} models is shown in Fig.~\ref{fig:age} and describes the evolution of the eclipsing stars. The primary has reached the Henyey track \citep{Henyey1965}, heating up and slowing down its contraction; whereas the secondary star continues contracting at roughly the same temperature along the Hayashi track \citep{Hayashi1961}. 
 The errors in the temperature ratio and sum of the radii predicted by the models come from the uncertainty in the measured masses (Table~\ref{table:eb}); 
the errors in the measured temperature ratio and sum of the radii were derived from the contour maps from the EB modeling (Fig.~\ref{fig:contours}). 
Assuming, as is standard for non-interacting, close binaries, that the eclipsing stars are coeval, Fig.~\ref{fig:age} shows that that the two ages derived independently from the theoretical evolutionary models 
compared to our measurements of the temperature ratio ($\sim$17 Myr; top panel) and the sum of the radii ($\sim$9 Myr; bottom panel) are not in mutual agreement.

\begin{figure*}
\centering
\includegraphics[width=1\hsize]{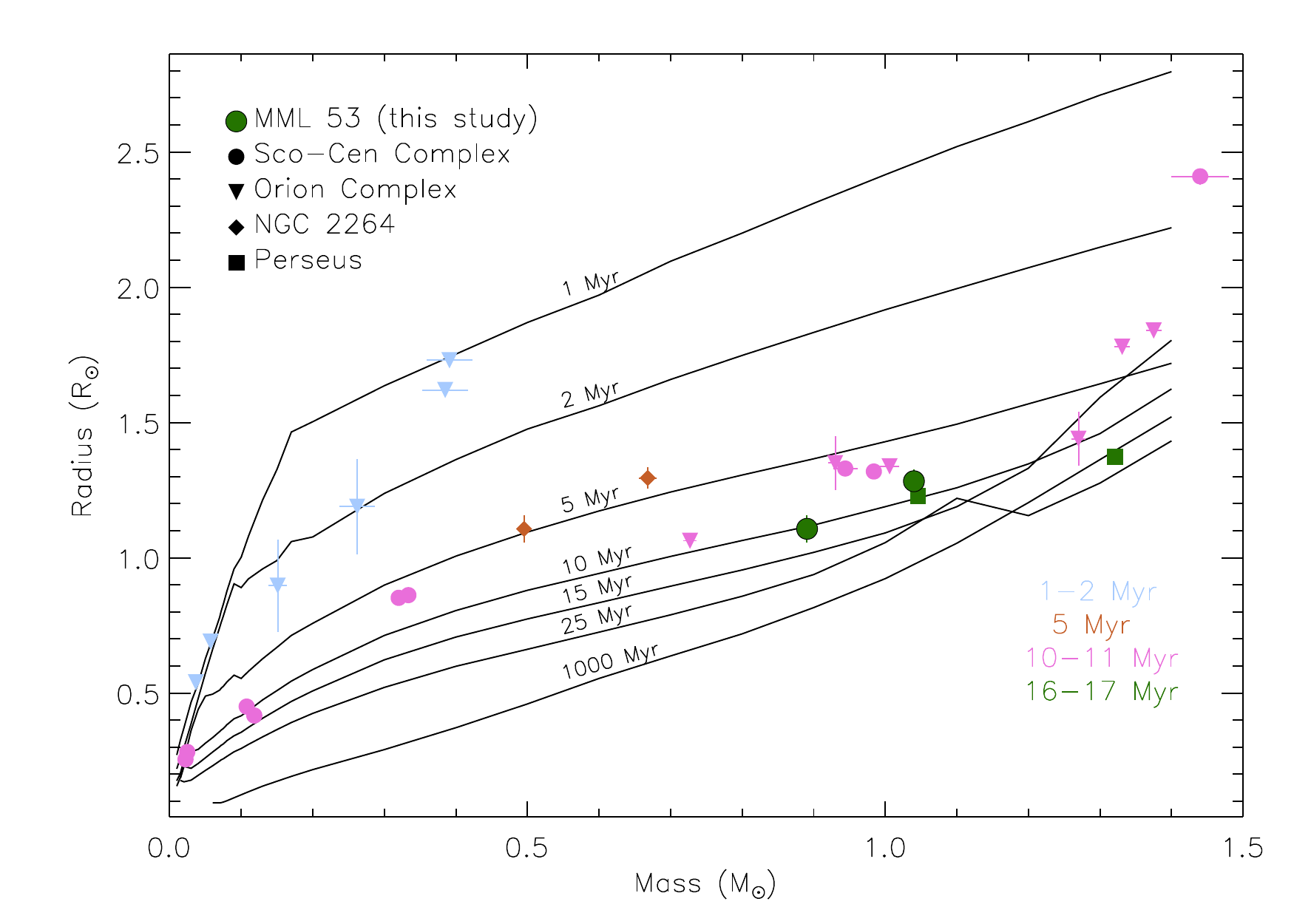}
\caption{Mass--radius diagram. We compare the known pre-main sequence stars in EBs with the evolutionary models of \citet{Baraffe2015}.  MML 53 is shown in the green dots with black edge. 
We also include the measurements of the known young EB stars (top-left legend) in: 
the Scorpius-Centaurus complex to which MML 53 belongs  
\citep[filled dots;][]{Alonso2015,David2016};
the Orion complex \citep[filled downward triangles;][]{Stempels2008,Covino2004,Torres2010,GomezMaqueoChew2012,Irwin2007,GomezMaqueoChew2009};
NGC 2264 
\citep[filled diamonds;][]{Gillen2014}, and
the Perseus complex \citep[filled squares;][]{Lacy2016}.
The color of each filled symbol represents the age of the system, as derived in previous analyses and are given in the bottom-right legend.   
The continuous black lines are the predicted radii of low-mass stars by the Baraffe models at different ages (from top to bottom: 1 Myr to 1 Gyr). All measurements are plotted with uncertainties; in the cases they are not visible, the uncertainties are smaller than the plotted symbols. 
}
\label{fig:mr}
\end{figure*}

\begin{figure*}[!h]
\centering
\includegraphics[width=1\hsize]{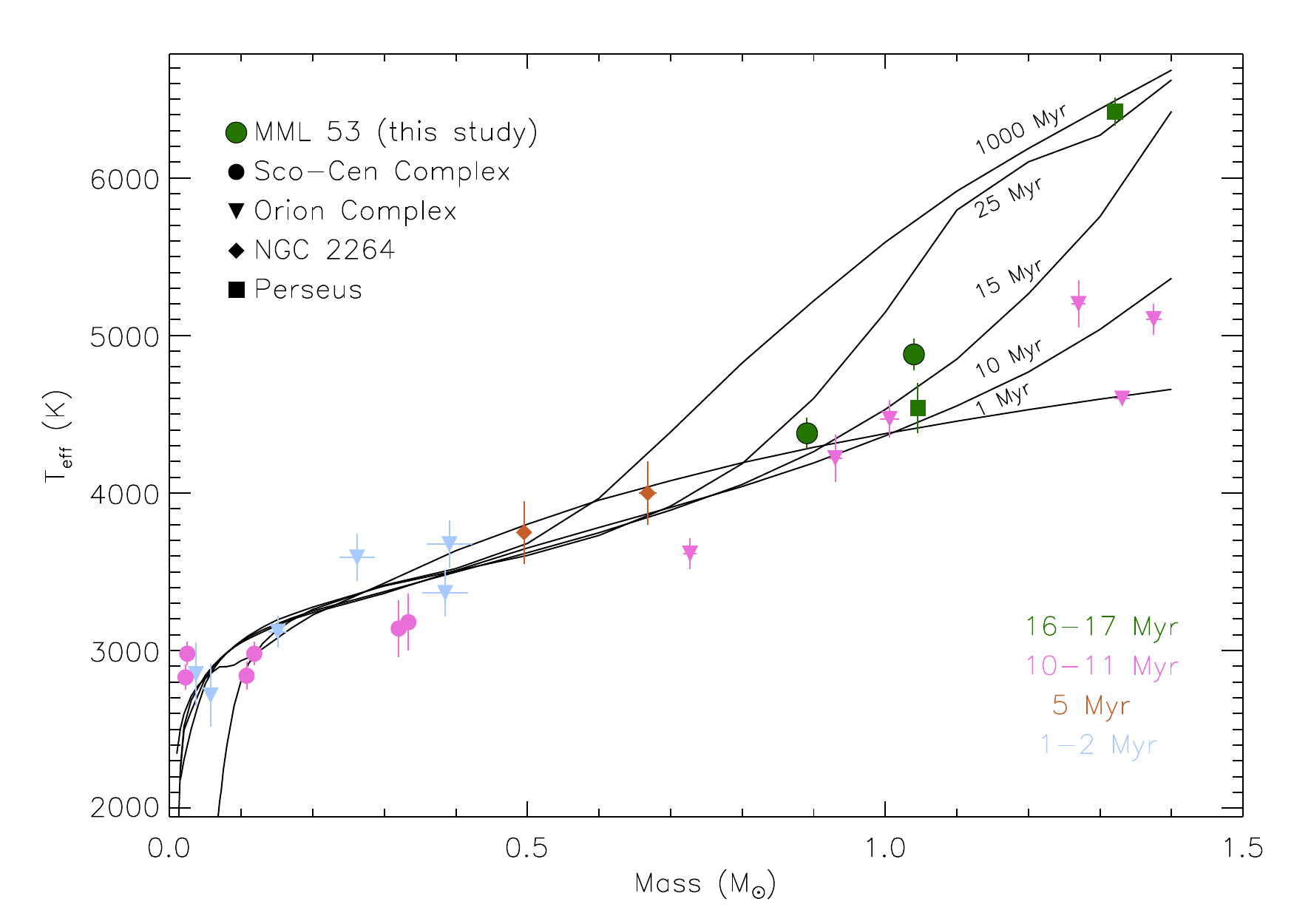}
\caption{Mass--temperature diagram. We show with continuous black lines the effective temperatures for low-mass stars ($<$ 1.4\msun) predicted
by the models of \citet{Baraffe2015}, as they evolve from 1 Myr to 1 Gyr. We compare the models to our measurements of the individual temperatures of the eclipsing components of MML~53 (green-filled circles with black edge), and the other young EBs in the literature with measured temperatures. The EBs, their symbols and colors are the same as in Fig.~\ref{fig:mr}. 
Each black line is an isochrone, and represents the expected temperatures of stars at a given age. 
} 
\label{fig:mt}
\end{figure*}
 
In-detail analyses \citep{Pecaut2016,Pecaut2012} present the mean age of the UCL subgroup (and of MML~53 itself) to be $16\pm2$~Myr based on a comparison between 14 high-mass, turn-off stars to the theoretical models for rotating stars from \citet{Ekstrom2012}, and on F-type and G-type members compared to pre-main sequence theoretical stellar evolution models \citep{Dotter2008,Baraffe2015,Tognelli2011,Chen2014}. 
An observed large spread in the ages of individual members of the subgroup is thought to be partly due to observational uncertainties and unresolved multiplicity, but also to an intrinsic spread of ages within the subgroup.  MML~53 is in the region of the association that is close to the average age, and is not in a part of the cluster that is thought to be much younger (high galactic longitudes) or older (edges of region). 
Additionally, the age derived from the temperature ratio (top panel; Fig.~\ref{fig:age}) is consistent with these age estimates for UCL;
thus we adopt an age of $16\pm2$~Myr for this specific object.  
Because the radii that we measured for the two eclipsing stars of MML~53 match to a much younger age (bottom panel in Fig.~\ref{fig:age}), the eclipsing components appear to be inflated. 

The presence of a third independent component in the spectrum of MML~53 had been identified from its discovery, as had the changes in the timing of the eclipses of the binary  \citep{Hebb2010}. However, it is only recently that the tertiary star has been resolved \citep{Schaefer2018} at its widest separation from the EB. It is in this paper that we have confirmed that the tertiary is a bound component of the MML~53 system and have constrained its mass dynamically. The $\sim$20\% uncertainty in the tertiary mass is a conservative estimate from the binary-tertiary model (\S\ref{sec:M3}), since the constraint from the optical spectra showing a lower luminosity for the tertiary has not been incorporated in the tertiary mass determination.

After carefully accounting for the third star, its effects on the measurement of the individual properties of the eclipsing components of MML~53 were minimized, allowing for a meaningful comparison between these measurements, the theoretical evolutionary models of \citet{Baraffe2015} and other direct measurements of pre-main sequence stars in double-lined, eclipsing binaries (see Figs.~\ref{fig:mr} and \ref{fig:mt}). 
The radii of both eclipsing components of MML~53 are  consistent with a single isochrone. However, they match the younger 10~Myr track (Fig.~\ref{fig:mr}). 
The radius of the primary star is inflated by 10\% with respect to the radius predicted by stellar models interpolated at 16 Myrs for a 1.04~\msun\ star, and the secondary radius is inflated by 15\% as predicted for a 0.89~\msun\ star.  

At 16~$\pm$ 2~Myr, MML~53 is very similar in age to the EB NP Per \citep[$\sim$17~Myr;][]{Lacy2016}, and importantly, 
the primary star of MML~53 
and the secondary of NP~Per 
have the same mass of 1.04 \msun, and as shown in the mass--radius diagram (Fig.~\ref{fig:mr}),
both stars have the same radius, measured independently, that appears to be too large, as predicted by the Baraffe models. 
More generally, the stars of a given eclipsing system, except for Par~1802 (blue, downward triangles at 0.4~\msun) and NP~Per (green squares), also fall on the same isochrone on the mass--radius diagram, showcasing that the theoretical models can describe the overall behavior of young, low-mass stars. However, the direct measurements of the EBs masses and radii appear to be younger than the ages derived by independent methods (e.g., from the study of the young associations: turn-off stars, mass of arrival at the main-sequence, lithium abundance). 
The apparent radius inflation of these young stars (including the eclipsing components of MML~53) could be due to physical processes not included in the evolutionary tracks, for example, the inhibition of convection due to magnetic fields that slow down contraction \citep{Feiden2016,MacDonald2017}, causing the lowest-mass young stars to appear younger. 

Stars that are younger than 10~Myr and have masses lower than 1.0~\msun\ have been at approximately the same temperature throughout their evolution, as they contract along the Hayashi track. This behavior is evident in the overlap of the theoretical isochrones and EB measurements shown in the mass--temperature diagram (Fig.~\ref{fig:mt}). In the case of MML~53 at 16~Myr and with masses straddling 1.0~\msun, its primary component has begun increasing in temperature,
while for the lower-mass secondary the tracks predict a small change in temperature ($\sim$100~K, within our errors) from 1 to 16~Myr. Importantly, our measurements of the individual temperatures follow the slope of the 16~Myr isochrone, even if slightly above it. 
Both stars are slightly hotter than predicted by the 16~Myr track. It is puzzling that the MML~53 eclipsing stars do not appear to be cooler, as would be expected if the inflated radii were due to magnetic activity, by inhibition of convection and/or spots \citep[e.g.,][]{Feiden2016,Somers2015}, in order for the stars to remain in thermal equilibrium.
In the mass--temperature plane, the primary of MML~53 and the secondary component of NP~Per are not consistent within one sigma, falling above and below the 16~Myr Baraffe model, respectively. Their individual temperatures would be consistent with each other, and with the 16~Myr model within two sigma.  
It is not surprising that in general the scatter in the mass--temperature diagram is larger than in the mass--radius diagram, as individual temperatures are harder to measure from the combined spectra of (at least) two stars. 

Some studies of triple systems composed of a close binary bound to an outer third star suggest that the formation of close binaries may be the result of tidal tightening of the inner binary due to the third star \citep[e.g.,][]{Tokovinin2006,Naoz2014}.  However at 16~Myr, the primary and secondary stars of MML~53 are already in a close binary ($\sim$0.03 au). Our results of MML~53 are more in agreement with recent population synthesis models suggesting that the mechanisms causing the tightening of the close  binary orbit occur in most cases early in the stars' evolution \citep{Moe2018}. 

\begin{acknowledgements}
    YGMC was supported in part by UNAM-PAPIIT IN-107518. 
    LHH was supported by grants NSF-AAG-1009810 and NSF-AAG-1312453. 
    We would like to thank G.\ Feiden,  R.\ Barnes and D.\ Fleming for useful discussions.
Based on observations collected at the European Organisation for Astronomical Research in the Southern Hemisphere under ESO program ID 
383.C-080.
      This research has made use of the services of the ESO Science Archive Facility. 
      This work makes use of observations from the LCOGT network.
      Based on observations at Cerro Tololo Inter-American Observatory, 
      which is operated by the Association of Universities for Research in Astronomy (AURA) under a cooperative agreement with the National Science Foundation. The CTIO observations made use of 1m and 1.5m telescopes operated by the SMARTS Consortium. 
WASP-South is hosted by the South African Astronomical Observatory and we are grateful for their ongoing support and assistance. Funding for WASP comes from
consortium universities and from the UK's Science and Technology Facilities
Council.

\end{acknowledgements}


\bibliographystyle{aa}
\bibliography{pmsebs}

\begin{thebibliography}{84}
\expandafter\ifx\csname natexlab\endcsname\relax\def\natexlab#1{#1}\fi

\bibitem[{Alonso {et~al.}(2015)Alonso, Deeg, Hoyer, Lodieu, Palle, \&
  Sanchis-Ojeda}]{Alonso2015}
Alonso, R., Deeg, H.~J., Hoyer, S., {et~al.} 2015, Astronomy {\&} Astrophysics,
  584, L8

\bibitem[{{Andersen}(1991)}]{Andersen1991}
{Andersen}, J. 1991, \aapr, 3, 91

\bibitem[{{Anderson} {et~al.}(2017){Anderson}, {Lai}, \&
  {Storch}}]{Anderson2017}
{Anderson}, K.~R., {Lai}, D., \& {Storch}, N.~I. 2017, \mnras, 467, 3066

\bibitem[{{Bagnuolo} \& {Gies}(1991)}]{Bagnuolo1991}
{Bagnuolo}, Jr., W.~G. \& {Gies}, D.~R. 1991, \apj, 376, 266

\bibitem[{{Baraffe} {et~al.}(2015){Baraffe}, {Homeier}, {Allard}, \&
  {Chabrier}}]{Baraffe2015}
{Baraffe}, I., {Homeier}, D., {Allard}, F., \& {Chabrier}, G. 2015, \aap, 577,
  A42

\bibitem[{{Bertin} \& {Arnouts}(1996)}]{Bertin1996}
{Bertin}, E. \& {Arnouts}, S. 1996, \aaps, 117, 393

\bibitem[{{Bouvier} \& {Bertout}(1989)}]{Bouvier1989}
{Bouvier}, J. \& {Bertout}, C. 1989, \aap, 211, 99

\bibitem[{{Cargile} {et~al.}(2008){Cargile}, {Stassun}, \&
  {Mathieu}}]{Cargile2008}
{Cargile}, P.~A., {Stassun}, K.~G., \& {Mathieu}, R.~D. 2008, \apj, 674, 329

\bibitem[{{Chen} {et~al.}(2014){Chen}, {Girardi}, {Bressan}, {Marigo},
  {Barbieri}, \& {Kong}}]{Chen2014}
{Chen}, Y., {Girardi}, L., {Bressan}, A., {et~al.} 2014, \mnras, 444, 2525

\bibitem[{{Collier Cameron} {et~al.}(2006){Collier Cameron}, {Pollacco},
  {Street}, {Lister}, {West}, {Wilson}, {Pont}, {Christian}, {Clarkson},
  {Enoch}, {Evans}, {Fitzsimmons}, {Haswell}, {Hellier}, {Hodgkin}, {Horne},
  {Irwin}, {Kane}, {Keenan}, {Norton}, {Parley}, {Osborne}, {Ryans}, {Skillen},
  \& {Wheatley}}]{CollierCameron2006}
{Collier Cameron}, A., {Pollacco}, D., {Street}, R.~A., {et~al.} 2006, \mnras,
  373, 799

\bibitem[{Covino {et~al.}(2004)Covino, Frasca, Alcal{\'{a}}, Paladino, \&
  Sterzik}]{Covino2004}
Covino, E., Frasca, A., Alcal{\'{a}}, J.~M., Paladino, R., \& Sterzik, M.~F.
  2004, Astronomy, 649, 637

\bibitem[{{Covino} {et~al.}(2001){Covino}, {Melo}, {Alcal{\'a}}, {Torres},
  {Fern{\'a}ndez}, {Frasca}, \& {Paladino}}]{Covino2001}
{Covino}, E., {Melo}, C., {Alcal{\'a}}, J.~M., {et~al.} 2001, \aap, 375, 130

\bibitem[{{David} {et~al.}(2016){David}, {Hillenbrand}, {Cody}, {Carpenter}, \&
  {Howard}}]{David2016}
{David}, T.~J., {Hillenbrand}, L.~A., {Cody}, A.~M., {Carpenter}, J.~M., \&
  {Howard}, A.~W. 2016, \apj, 816, 21

\bibitem[{{de Zeeuw} {et~al.}(1999){de Zeeuw}, {Hoogerwerf}, {de Bruijne},
  {Brown}, \& {Blaauw}}]{deZeeuw1999}
{de Zeeuw}, P.~T., {Hoogerwerf}, R., {de Bruijne}, J.~H.~J., {Brown}, A.~G.~A.,
  \& {Blaauw}, A. 1999, \aj, 117, 354

\bibitem[{{Donati} {et~al.}(1997){Donati}, {Semel}, {Carter}, {Rees}, \&
  {Collier Cameron}}]{Donati1997}
{Donati}, J.-F., {Semel}, M., {Carter}, B.~D., {Rees}, D.~E., \& {Collier
  Cameron}, A. 1997, \mnras, 291, 658

\bibitem[{{Dotter} {et~al.}(2008){Dotter}, {Chaboyer}, {Jevremovi{\'c}},
  {Kostov}, {Baron}, \& {Ferguson}}]{Dotter2008}
{Dotter}, A., {Chaboyer}, B., {Jevremovi{\'c}}, D., {et~al.} 2008, \apjs, 178,
  89

\bibitem[{{Eastman} {et~al.}(2013){Eastman}, {Gaudi}, \& {Agol}}]{Eastman2013}
{Eastman}, J., {Gaudi}, B.~S., \& {Agol}, E. 2013, \pasp, 125, 83

\bibitem[{{Eastman} {et~al.}(2010){Eastman}, {Siverd}, \& {Gaudi}}]{BJDTDB}
{Eastman}, J., {Siverd}, R., \& {Gaudi}, B.~S. 2010, \pasp, 122, 935

\bibitem[{{Ekstr{\"o}m} {et~al.}(2012){Ekstr{\"o}m}, {Georgy}, {Eggenberger},
  {Meynet}, {Mowlavi}, {Wyttenbach}, {Granada}, {Decressin}, {Hirschi},
  {Frischknecht}, {Charbonnel}, \& {Maeder}}]{Ekstrom2012}
{Ekstr{\"o}m}, S., {Georgy}, C., {Eggenberger}, P., {et~al.} 2012, \aap, 537,
  A146

\bibitem[{{Feiden}(2016)}]{Feiden2016}
{Feiden}, G.~A. 2016, \aap, 593, A99

\bibitem[{{Findeisen} \& {Hillenbrand}(2010)}]{Findeisen2010}
{Findeisen}, K. \& {Hillenbrand}, L. 2010, \aj, 139, 1338

\bibitem[{{Foreman-Mackey} {et~al.}(2013){Foreman-Mackey}, {Hogg}, {Lang}, \&
  {Goodman}}]{ForemanMackey2013}
{Foreman-Mackey}, D., {Hogg}, D.~W., {Lang}, D., \& {Goodman}, J. 2013, \pasp,
  125, 306

\bibitem[{{Gillen} {et~al.}(2014){Gillen}, {Aigrain}, {McQuillan}, {Bouvier},
  {Hodgkin}, {Alencar}, {Terquem}, {Southworth}, {Gibson}, {Cody}, {Lendl},
  {Morales-Calder{\'o}n}, {Favata}, {Stauffer}, \& {Micela}}]{Gillen2014}
{Gillen}, E., {Aigrain}, S., {McQuillan}, A., {et~al.} 2014, \aap, 562, A50

\bibitem[{{Gillen} {et~al.}(2017){Gillen}, {Aigrain}, {Terquem}, {Bouvier},
  {Alencar}, {Gandolfi}, {Stauffer}, {Cody}, {Venuti}, {Almeida}, {Micela},
  {Favata}, \& {Deeg}}]{Gillen2017}
{Gillen}, E., {Aigrain}, S., {Terquem}, C., {et~al.} 2017, \aap, 599, A27

\bibitem[{{G{\'{o}}mez Maqueo Chew} {et~al.}(2009){G{\'{o}}mez Maqueo Chew},
  Stassun, Pr{\v{s}}a, \& Mathieu}]{GomezMaqueoChew2009}
{G{\'{o}}mez Maqueo Chew}, Y., Stassun, K.~G., Pr{\v{s}}a, A., \& Mathieu,
  R.~D. 2009, The Astrophysical Journal, 699, 1196

\bibitem[{{G{\'{o}}mez Maqueo Chew} {et~al.}(2012){G{\'{o}}mez Maqueo Chew},
  Stassun, Pr{\v{s}}a, Stempels, Hebb, Barnes, Heller, \&
  Mathieu}]{GomezMaqueoChew2012}
{G{\'{o}}mez Maqueo Chew}, Y., Stassun, K.~G., Pr{\v{s}}a, A., {et~al.} 2012,
  The Astrophysical Journal, 745, 58

\bibitem[{{Gustafsson} {et~al.}(2008){Gustafsson}, {Edvardsson}, {Eriksson},
  {J{\o}rgensen}, {Nordlund}, \& {Plez}}]{Gustafsson2008}
{Gustafsson}, B., {Edvardsson}, B., {Eriksson}, K., {et~al.} 2008, \aap, 486,
  951

\bibitem[{{Hayashi}(1961)}]{Hayashi1961}
{Hayashi}, C. 1961, \pasj, 13

\bibitem[{{Hebb} {et~al.}(2011){Hebb}, {Cegla}, {Stassun}, {Stempels},
  {Cargile}, \& {Palladino}}]{Hebb2011}
{Hebb}, L., {Cegla}, H.~M., {Stassun}, K.~G., {et~al.} 2011, \aap, 531, A61

\bibitem[{{Hebb} {et~al.}(2010){Hebb}, {Stempels}, {Aigrain},
  {Collier-Cameron}, {Hodgkin}, {Irwin}, {Maxted}, {Pollacco}, {Street},
  {Wilson}, \& {Stassun}}]{Hebb2010}
{Hebb}, L., {Stempels}, H.~C., {Aigrain}, S., {et~al.} 2010, \aap, 522, A37

\bibitem[{{Henyey} {et~al.}(1965){Henyey}, {Vardya}, \&
  {Bodenheimer}}]{Henyey1965}
{Henyey}, L., {Vardya}, M.~S., \& {Bodenheimer}, P. 1965, \apj, 142, 841

\bibitem[{{Irwin} {et~al.}(2007){Irwin}, {Aigrain}, {Hodgkin}, {Stassun},
  {Hebb}, {Irwin}, {Moraux}, {Bouvier}, {Alapini}, {Alexander}, {Bramich},
  {Holtzman}, {Mart{\'{\i}}n}, {McCaughrean}, {Pont}, {Verrier}, \& {Zapatero
  Osorio}}]{Irwin2007}
{Irwin}, J., {Aigrain}, S., {Hodgkin}, S., {et~al.} 2007, \mnras, 380, 541

\bibitem[{{Irwin} \& {Lewis}(2001)}]{IrwinLewis2001}
{Irwin}, M. \& {Lewis}, J. 2001, \nar, 45, 105

\bibitem[{{Kallrath} \& {Milone}(2009)}]{Kallrath2009}
{Kallrath}, J. \& {Milone}, E.~F. 2009, {Eclipsing Binary Stars: Modeling and
  Analysis}

\bibitem[{{Kiraga}(2012)}]{Kiraga2012}
{Kiraga}, M. 2012, \actaa, 62, 67

\bibitem[{{Kochukhov} {et~al.}(2010){Kochukhov}, {Makaganiuk}, \&
  {Piskunov}}]{Kochukhov2010}
{Kochukhov}, O., {Makaganiuk}, V., \& {Piskunov}, N. 2010, \aap, 524, A5

\bibitem[{{Kopal}(1959)}]{Kopal1959}
{Kopal}, Z. 1959, {Close binary systems}

\bibitem[{{Kraus} {et~al.}(2015){Kraus}, {Cody}, {Covey}, {Rizzuto}, {Mann}, \&
  {Ireland}}]{Kraus2015}
{Kraus}, A.~L., {Cody}, A.~M., {Covey}, K.~R., {et~al.} 2015, \apj, 807, 3

\bibitem[{{Kupka} {et~al.}(1999){Kupka}, {Piskunov}, {Ryabchikova}, {Stempels},
  \& {Weiss}}]{Kupka1999}
{Kupka}, F., {Piskunov}, N., {Ryabchikova}, T.~A., {Stempels}, H.~C., \&
  {Weiss}, W.~W. 1999, \aaps, 138, 119

\bibitem[{{Lacy} {et~al.}(2016){Lacy}, {Fekel}, {Pavlovski}, {Torres}, \&
  {Muterspaugh}}]{Lacy2016}
{Lacy}, C.~H.~S., {Fekel}, F.~C., {Pavlovski}, K., {Torres}, G., \&
  {Muterspaugh}, M.~W. 2016, \aj, 152, 2

\bibitem[{{Lodieu} {et~al.}(2015){Lodieu}, {Alonso}, {Gonz{\'a}lez
  Hern{\'a}ndez}, {Sanchis-Ojeda}, {Narita}, {Kawashima}, {Kawauchi},
  {Su{\'a}rez Mascare{\~n}o}, {Deeg}, {Prieto Arranz}, {Rebolo}, {Pall{\'e}},
  {B{\'e}jar}, {Ferragamo}, \& {Rubi{\~n}o-Mart{\'{\i}}n}}]{Lodieu2015}
{Lodieu}, N., {Alonso}, R., {Gonz{\'a}lez Hern{\'a}ndez}, J.~I., {et~al.} 2015,
  \aap, 584, A128

\bibitem[{{Lucy} \& {Sweeney}(1971)}]{Lucy1971}
{Lucy}, L.~B. \& {Sweeney}, M.~A. 1971, \aj, 76, 544

\bibitem[{{MacDonald} \& {Mullan}(2017)}]{MacDonald2017}
{MacDonald}, J. \& {Mullan}, D.~J. 2017, \apj, 834, 67

\bibitem[{Mamajek {et~al.}(2002)Mamajek, Meyer, \& Liebert}]{Mamajek2002}
Mamajek, E., Meyer, M., \& Liebert, J. 2002, $\backslash$Aj, 124, 1670

\bibitem[{{Mathieu} {et~al.}(2007){Mathieu}, {Baraffe}, {Simon}, {Stassun}, \&
  {White}}]{Mathieu2007}
{Mathieu}, R.~D., {Baraffe}, I., {Simon}, M., {Stassun}, K.~G., \& {White}, R.
  2007, Protostars and Planets V, 411

\bibitem[{{Mazeh}(2008)}]{Mazeh2008}
{Mazeh}, T. 2008, in EAS Publications Series, Vol.~29, EAS Publications Series,
  ed. M.-J. {Goupil} \& J.-P. {Zahn}, 1--65

\bibitem[{{Moe} \& {Kratter}(2018)}]{Moe2018}
{Moe}, M. \& {Kratter}, K.~M. 2018, \apj, 854, 44

\bibitem[{{Morales} {et~al.}(2010){Morales}, {Gallardo}, {Ribas}, {Jordi},
  {Baraffe}, \& {Chabrier}}]{Morales2010}
{Morales}, J.~C., {Gallardo}, J., {Ribas}, I., {et~al.} 2010, \apj, 718, 502

\bibitem[{{Morales-Calder{\'o}n} {et~al.}(2012){Morales-Calder{\'o}n},
  {Stauffer}, {Stassun}, {Vrba}, {Prato}, {Hillenbrand}, {Terebey}, {Covey},
  {Rebull}, {Terndrup}, {Gutermuth}, {Song}, {Plavchan}, {Carpenter},
  {Marchis}, {Garc{\'{\i}}a}, {Margheim}, {Luhman}, {Angione}, \&
  {Irwin}}]{Morales2012}
{Morales-Calder{\'o}n}, M., {Stauffer}, J.~R., {Stassun}, K.~G., {et~al.} 2012,
  \apj, 753, 149

\bibitem[{{Naoz} \& {Fabrycky}(2014)}]{Naoz2014}
{Naoz}, S. \& {Fabrycky}, D.~C. 2014, \apj, 793, 137

\bibitem[{{Nidever} {et~al.}(2002){Nidever}, {Marcy}, {Butler}, {Fischer}, \&
  {Vogt}}]{Nidever2002}
{Nidever}, D.~L., {Marcy}, G.~W., {Butler}, R.~P., {Fischer}, D.~A., \& {Vogt},
  S.~S. 2002, \apjs, 141, 503

\bibitem[{{Padgett}(1996)}]{Padgett1996}
{Padgett}, D.~L. 1996, \apj, 471, 847

\bibitem[{{Pecaut} \& {Mamajek}(2016)}]{Pecaut2016}
{Pecaut}, M.~J. \& {Mamajek}, E.~E. 2016, \mnras, 461, 794

\bibitem[{Pecaut {et~al.}(2012)Pecaut, Mamajek, \& Bubar}]{Pecaut2012}
Pecaut, M.~J., Mamajek, E.~E., \& Bubar, E.~J. 2012, The Astrophysical Journal,
  746, 154

\bibitem[{{Perova} {et~al.}(1966){Perova}, {Ureche}, \&
  {Kholopov}}]{Perova1966}
{Perova}, N.~B., {Ureche}, V., \& {Kholopov}, P.~N. 1966, Astronomicheskij
  Tsirkulyar, 367, 3

\bibitem[{{Piskunov} \& {Valenti}(2017)}]{Piskunov2017}
{Piskunov}, N. \& {Valenti}, J.~A. 2017, \aap, 597, A16

\bibitem[{{Piskunov} {et~al.}(1995){Piskunov}, {Kupka}, {Ryabchikova}, {Weiss},
  \& {Jeffery}}]{Piskunov1995}
{Piskunov}, N.~E., {Kupka}, F., {Ryabchikova}, T.~A., {Weiss}, W.~W., \&
  {Jeffery}, C.~S. 1995, \aaps, 112, 525

\bibitem[{{Piskunov} \& {Valenti}(2002)}]{Piskunov2002}
{Piskunov}, N.~E. \& {Valenti}, J.~A. 2002, \aap, 385, 1095

\bibitem[{{Pollacco} {et~al.}(2006){Pollacco}, {Skillen}, {Collier Cameron},
  {Christian}, {Hellier}, {Irwin}, {Lister}, {Street}, {West}, {Anderson},
  {Clarkson}, {Deeg}, {Enoch}, {Evans}, {Fitzsimmons}, {Haswell}, {Hodgkin},
  {Horne}, {Kane}, {Keenan}, {Maxted}, {Norton}, {Osborne}, {Parley}, {Ryans},
  {Smalley}, {Wheatley}, \& {Wilson}}]{Pollacco2006}
{Pollacco}, D.~L., {Skillen}, I., {Collier Cameron}, A., {et~al.} 2006, \pasp,
  118, 1407

\bibitem[{{Popper} \& {Etzel}(1981)}]{popperetzel}
{Popper}, D.~M. \& {Etzel}, P.~B. 1981, \aj, 86, 102

\bibitem[{{Pr{\v s}a} {et~al.}(2016){Pr{\v s}a}, {Harmanec}, {Torres},
  {Mamajek}, {Asplund}, {Capitaine}, {Christensen-Dalsgaard}, {Depagne},
  {Haberreiter}, {Hekker}, {Hilton}, {Kopp}, {Kostov}, {Kurtz}, {Laskar},
  {Mason}, {Milone}, {Montgomery}, {Richards}, {Schmutz}, {Schou}, \&
  {Stewart}}]{Prsa2016}
{Pr{\v s}a}, A., {Harmanec}, P., {Torres}, G., {et~al.} 2016, \aj, 152, 41

\bibitem[{{Pr{\v s}a} \& {Zwitter}(2005)}]{Prsa2005}
{Pr{\v s}a}, A. \& {Zwitter}, T. 2005, \apj, 628, 426

\bibitem[{{Schaefer} {et~al.}(2018){Schaefer}, {Prato}, \&
  {Simon}}]{Schaefer2018}
{Schaefer}, G.~H., {Prato}, L., \& {Simon}, M. 2018, \aj, 155, 109

\bibitem[{{Somers} \& {Pinsonneault}(2015)}]{Somers2015}
{Somers}, G. \& {Pinsonneault}, M.~H. 2015, \apj, 807, 174

\bibitem[{{Southworth} {et~al.}(2007){Southworth}, {Bruntt}, \&
  {Buzasi}}]{southworth2007}
{Southworth}, J., {Bruntt}, H., \& {Buzasi}, D.~L. 2007, \aap, 467, 1215

\bibitem[{{Stassun} {et~al.}(2014){Stassun}, {Feiden}, \&
  {Torres}}]{Stassun2014}
{Stassun}, K.~G., {Feiden}, G.~A., \& {Torres}, G. 2014, \nar, 60, 1

\bibitem[{{Stassun} {et~al.}(2008){Stassun}, {Mathieu}, {Cargile}, {Aarnio},
  {Stempels}, \& {Geller}}]{Stassun2008}
{Stassun}, K.~G., {Mathieu}, R.~D., {Cargile}, P.~A., {et~al.} 2008, \nat, 453,
  1079

\bibitem[{{Stassun} {et~al.}(2006){Stassun}, {Mathieu}, \&
  {Valenti}}]{Stassun2006}
{Stassun}, K.~G., {Mathieu}, R.~D., \& {Valenti}, J.~A. 2006, \nat, 440, 311

\bibitem[{{Stassun} {et~al.}(2007){Stassun}, {Mathieu}, \&
  {Valenti}}]{Stassun2007}
{Stassun}, K.~G., {Mathieu}, R.~D., \& {Valenti}, J.~A. 2007, \apj, 664, 1154

\bibitem[{{Stassun} {et~al.}(2004){Stassun}, {Mathieu}, {Vaz}, {Stroud}, \&
  {Vrba}}]{Stassun2004}
{Stassun}, K.~G., {Mathieu}, R.~D., {Vaz}, L.~P.~R., {Stroud}, N., \& {Vrba},
  F.~J. 2004, \apjs, 151, 357

\bibitem[{{Stempels} \& {Hebb}(2011)}]{Stempels2011}
{Stempels}, H.~C. \& {Hebb}, L. 2011, in Astronomical Society of the Pacific
  Conference Series, Vol. 448, 16th Cambridge Workshop on Cool Stars, Stellar
  Systems, and the Sun, ed. C.~{Johns-Krull}, M.~K. {Browning}, \& A.~A.
  {West}, 747

\bibitem[{{Stempels} {et~al.}(2008){Stempels}, {Hebb}, {Stassun}, {Holtzman},
  {Dunstone}, {Glowienka}, \& {Frandsen}}]{Stempels2008}
{Stempels}, H.~C., {Hebb}, L., {Stassun}, K.~G., {et~al.} 2008, \aap, 481, 747

\bibitem[{{Tody}(1993)}]{Tody93}
{Tody}, D. 1993, in Astronomical Society of the Pacific Conference Series,
  Vol.~52, Astronomical Data Analysis Software and Systems II, ed. R.~J.
  {Hanisch}, R.~J.~V. {Brissenden}, \& J.~{Barnes}, 173

\bibitem[{{Tognelli} {et~al.}(2011){Tognelli}, {Prada Moroni}, \&
  {Degl'Innocenti}}]{Tognelli2011}
{Tognelli}, E., {Prada Moroni}, P.~G., \& {Degl'Innocenti}, S. 2011, \aap, 533,
  A109

\bibitem[{{Tokovinin} {et~al.}(2006){Tokovinin}, {Thomas}, {Sterzik}, \&
  {Udry}}]{Tokovinin2006}
{Tokovinin}, A., {Thomas}, S., {Sterzik}, M., \& {Udry}, S. 2006, \aap, 450,
  681

\bibitem[{Torres {et~al.}(2006)Torres, Quast, da~Silva, de~la Reza, Melo, \&
  Sterzik}]{Torres2006}
Torres, C. A.~O., Quast, G.~R., da~Silva, L., {et~al.} 2006, Astronomy and
  Astrophysics, 460, 695

\bibitem[{{Torres} {et~al.}(2010){Torres}, {Andersen}, \&
  {Gim{\'e}nez}}]{Torres2010}
{Torres}, G., {Andersen}, J., \& {Gim{\'e}nez}, A. 2010, \aapr, 18, 67

\bibitem[{{Valenti} \& {Piskunov}(1996)}]{Valenti1996}
{Valenti}, J.~A. \& {Piskunov}, N. 1996, \aaps, 118, 595

\bibitem[{{van Eyken} {et~al.}(2011){van Eyken}, {Ciardi}, {Rebull},
  {Stauffer}, {Akeson}, {Beichman}, {Boden}, {von Braun}, {Gelino}, {Hoard},
  {Howell}, {Kane}, {Plavchan}, {Ram{\'{\i}}rez}, {Bloom}, {Cenko}, {Kasliwal},
  {Kulkarni}, {Law}, {Nugent}, {Ofek}, {Poznanski}, {Quimby}, {Grillmair},
  {Laher}, {Levitan}, {Mattingly}, \& {Surace}}]{vanEyken2011}
{van Eyken}, J.~C., {Ciardi}, D.~R., {Rebull}, L.~M., {et~al.} 2011, \aj, 142,
  60

\bibitem[{White {et~al.}(2007)White, Gabor, \& Hillenbrand}]{White2007}
White, R.~J., Gabor, J.~M., \& Hillenbrand, L.~a. 2007, The Astronomical
  Journal, 133, 2524

\bibitem[{Wichmann {et~al.}(1997)Wichmann, Sterzik, Krautter, Metanomski, \&
  Voges}]{Wichmann1997a}
Wichmann, R., Sterzik, M., Krautter, J., Metanomski, A., \& Voges, W. 1997,
  Astronomy and Astrophysics, 326, 211

\bibitem[{{Wilson}(1990)}]{Wilson1990}
{Wilson}, R.~E. 1990, \apj, 356, 613

\bibitem[{{Windmiller} {et~al.}(2010){Windmiller}, {Orosz}, \&
  {Etzel}}]{Windmiller2010}
{Windmiller}, G., {Orosz}, J.~A., \& {Etzel}, P.~B. 2010, \apj, 712, 1003

\bibitem[{{Zahn}(1977)}]{Zahn1977}
{Zahn}, J.-P. 1977, \aap, 57, 383

\end{thebibliography}

\end{document}